\documentclass{revtex4}
\usepackage{epsfig}
\usepackage[dvips]{color}
\def\bea{\begin{eqnarray}}
\def\eea{\end{eqnarray}}
\def\bec{\begin{center}}
\def\ec{\end{center}}

\def\beq{\begin{equation}}
\def\eeq{\end{equation}}

\newcommand\lsim{\mathrel{\rlap{\lower4pt\hbox{\hskip1pt$\sim$}}
    \raise1pt\hbox{$<$}}}
\newcommand\gsim{\mathrel{\rlap{\lower4pt\hbox{\hskip1pt$\sim$}}
    \raise1pt\hbox{$>$}}}

\newcommand\cb{\textcolor{cyan}}
\newcommand\mhalf{M_{1/2}}

\begin{document}
\draft
%\tighten
\preprint{KAIST-TH 04/09}
\title{\large \bf
 Scherk-Schwarz
Supersymmetry Breaking for Quasi-localized Matter Fields \\
and Supersymmetry Flavor Violation
}
\author{Hiroyuki Abe\footnote{abe@hep.kaist.ac.kr},
Kiwoon Choi\footnote{kchoi@hep.kaist.ac.kr},
Kwang-Sik Jeong\footnote{ksjeong@hep.kaist.ac.kr},
Ken-ichi Okumura\footnote{okumura@hep.kaist.ac.kr}}
\address{Department of Physics, Korea Advanced Institute of
Science and Technology\\ Daejeon
305-701, Korea}
\date{\today}
%\maketitle
%%%%%%%%%%%%%%%%%%%%%%%%%%%%%%%%%%%%%%%%%%%%%%%%%%%%%%
\begin{abstract}
%%%%%%%%%%%%%%%%%%%%%%%%%%%%%%%%%%%%%%%%%%%%%%%%%%%%%%
We examine the soft supersymmetry breaking parameters
induced by the Scherk-Schwarz (SS) boundary
condition in 5-dimensional orbifold field theory in which
the quark and lepton zero modes are quasi-localized at
the orbifold fixed points to generate the hierarchical
Yukawa couplings.
In such theories, the radion corresponds to
a flavon to generate the flavor hierarchy and at the same time
plays the role of the messenger of supersymmetry breaking.
As a consequence, the resulting soft scalar masses and
trilinear $A$-parameters of matter zero modes
at the compactification scale are highly flavor-dependent,
thereby can lead to dangerous flavor violations at low energy scales.
We analyze in detail the low energy flavor violations
in SS-dominated supersymmetry breaking scenario
under the assumption that the compactification scale
is close to the grand unification scale and the 4-dimensional effective
theory below the compactification scale is given by the minimal
supersymmetric standard model.
Our analysis can be applied  to any supersymmetry breaking
mechanism giving a sizable $F$-component of the radion superfield,
e.g. the hidden gaugino condensation model.

%%%%%%%%%%%%%%%%%%%%%%%%%%%%%%%%%%%%%%%%%%%%%%%%%%%%%%%
\end{abstract}
%%%%%%%%%%%%%%%%%%%%%%%%%%%%%%%%%%%%%%%%%%%%%%%%%%%%%%%
\maketitle

\section{Introduction}

Supersymmetry (SUSY) is one of the prime candidates for
new physics beyond the standard model (SM) \cite{Nilles:1983ge}.
An important issue in supersymmetric theories is to
understand how SUSY is broken in low energy
world. It has been known that theories with compact extra dimension
provide an attractive way to break SUSY by
imposing nontrivial boundary conditions on the field variables.
This mechanism which has been proposed originally by Scherk and Schwarz (SS)
\cite{Scherk:1978ta} can be interpreted as a SUSY breaking
induced by the auxiliary component of higher dimensional supergravity
(SUGRA) multiplet \cite{Kaplan:2001cg}.
Extra dimension can provide
also an attractive mechanism to
generate hierarchical Yukawa
couplings \cite{Arkani-Hamed:1999dc}.
The quark and lepton fields can be quasi-localized
in extra dimension, and then
their 4-dimensional (4D) Yukawa couplings involve
the wavefunction overlap factor
$e^{-M\pi R}$ where
$M$ is a combination of mass parameters in higher dimensional
theory and $R$ is the size of extra dimension.
This would result in hierarchically different Yukawa couplings
even when the fundamental mass parameters have
the same order of magnitudes.

A simple and natural theoretical framework
for the quasi-localization of matter zero modes
is supersymmetric 5D orbifold field theory.
If a matter hypermultiplet in 5D orbifold SUGRA
has a non-zero gauge charge for the graviphoton
and/or for an ordinary $U(1)_{FI}$ vector multiplet which
has non-zero boundary Fayet-Iliopoulos (FI) terms \cite{Barbieri:2002ic},
it obtains a non-zero 5D kink mass
$M\epsilon(y)$
where $\epsilon(y)=\pm 1$ is the periodic sign function
on $S^1/Z_2$ whose fundamental domain is given by
$0\leq y\leq \pi$.
In the presence of such kink mass,
the matter zero mode  becomes quasi-localized
at one of the orbifold fixed points $y=0,\pi$
with a wavefunction given by $e^{-MR|y|}$.
A 5D orbifold SUGRA provides also a simple theoretical framework
for the SS SUSY breaking.
The theory admits a continuous twist of $SU(2)_R$ boundary condition
under the discrete shift $y\rightarrow y+2\pi$ which would
break the $N=1$ SUSY survived from the $Z_2$-orbifolding \cite{Kaplan:2001cg}.

In this paper, we wish to examine some physical consequences
of implementing the quasi-localization of matter zero modes
and the SS SUSY breaking
simultaneously within 5D orbifold field theories,
particularly the flavor structure of soft parameters and the resulting
low energy flavor violations.
In section 2,
 we first discuss some features of
5D orbifold SUGRA related to the quasi-localization
of matter fields and  also
the SS SUSY breaking.
We then compute the soft parameters of quasi-localized matter fields
induced by the SS boundary condition in generic
5D orbifold SUGRA.
We show explicitly that the zero mode soft parameters from
the SS boundary condition are same as the ones
induced by the radion $F$-component in 4D effective SUGRA,
and thus our analysis applies to any SUSY breaking mechanism
giving a sizable $F$-component of the radion superfield \cite{radionmediation},
e.g. the hidden gaugino condensation model.
This means that the radion superfield which corresponds to
a flavon for the Yukawa hierarchy
plays the role of the messenger of SUSY breaking.
As a consequence, the resulting soft scalar masses and
trilinear $A$ parameters of matter zero modes
at the compactification scale are highly flavor-dependent,
thereby can lead to dangerous flavor violation at low energy scales.
In particular, the predicted shape of soft parameters at the compactification scale
indicates that the compactification scale should be much higher
than the weak scale in order for the model to be
phenomenologically viable.

In 5D orbifold SUGRA, 5D kink masses $M_I\epsilon(y)$ responsible for quasi-localization
have quantized-values if the graviphoton and/or $U(1)_{FI}$ gauge charges are quantized.
An important feature of the SS SUSY breaking is that,
if the kink masses are quantized,
the resulting soft scalar masses and the trilinear scalar couplngs (divided by the corresponding
Yukawa couplings) at the compactification scale are quantized also in the leading approximation.
This feature provides a natural mechanism to suppress dangerous flavor violations
since the flavor violating amplitudes appear in a form
$f(M_I)-f(M_J)$, thus are canceled
when some of the quantized kink masses are degenerate.

In section 3, we analyze in detail the resulting
low energy flavor violations under the assumption
that the compactification scale $M_c$ is close to the grand unification
scale $M_{GUT}\sim 2\times 10^{16}$ GeV and the 4D effective theory below
$M_c$ is given by the minimal supersymmetric standard model (MSSM).
We find that many of the low energy flavor violations are appropriately
suppressed,
however generically $\epsilon_K$ and $\mu\rightarrow e\gamma$
can be too large if the SS boundary condition is the major source of
SUSY breaking.
As summarized in Table I,
if either the $SU(2)_L$-doublet lepton kink masses or
the $SU(2)_L$-singlet lepton kink masses are flavor-independent,
the $\mu\rightarrow e\gamma$ bound can be satisfied
for a reasonable range of the involved continuous parameters.
Tables II$-$III contain the predictions for
lepton flavor violating processes
 of the models which can satisfy the $\mu\to e\gamma$ constraint
with a mild tuning of the involved parameters.
Different choices of the lepton kink masses
predict different patterns of lepton flavor violations.
In particular, when the $SU(2)_L$-doublet lepton kink masses are degenerate,
the predicted chirality structure of decay modes
is opposite to the other case with degenerate $SU(2)_L$-singlet lepton kink masses
which has the same chirality structure as
the lepton flavor violating decays in seesaw models \cite{seesaw,Hisano:1998fj,muegpol}.
For $\epsilon_K$, it is more difficult to make the SUSY contribution
small enough since the model is constrained to yield the correct CKM mixing angles
as well as the correct quark mass hierarchy. Again a possible option is
that the kink masses of the $SU(2)_L$ singlet down quarks
are flavor-independent. However in this case, in order to
produce the correct quark mass eigenvalues and CKM mixing angles,
one needs to assume that some
boundary Yukawa couplings are abnormally large (or small) by a factor of $4\sim 5$
($0.2\sim 0.3$) compared to the values suggested by the naive dimensional
analysis.
Table IV summarizes the SUSY contributions to $\epsilon_K$ for some choices of
the quark kink masses.
We finally discuss the SUSY contributions
to other flavor-violating amplitudes
in SS SUSY breaking scenario, e.g.
the $b\to s\gamma$ rate, $\epsilon'/\epsilon_K$ and
the $K^0$-$\bar{K}^0$ and $B^0$-$\bar{B}^0$ mass differences,
which turn out to be either well below
or at most comparable to the SM contributions.
%Table V summarizes the SUSY contributions to $\Delta M_K$ and
%$\Delta M_{B_d,B_s}$ for the models of Table IV.

\section{Supersymmetry breaking by boundary condition for quasi-localized
matter fields in 5D orbifold supergravity}

In this section, we first discuss some features of
5D orbifold SUGRA related to the quasi-localization
of matter fields and  also
the SS SUSY breaking by boundary condition.
We then compute the soft
parameters of quasi-localized matter fields
induced by the SS boundary condition and compare the results
with the radion-mediated soft parameters
in 4D effective SUGRA.

Let us consider
a generic SUGRA-coupled 5D gauge theory on
$S^1/Z_2$.
The action of the theory is given by
\cite{Altendorfer:2000rr}
\bea \label{5daction1}
S &=& \int d^5x\sqrt{-G}\,\left[\,\frac{1}{2}\left(\, {\cal
R}+\bar{\Psi}^i_M\gamma^{MNP}D_N\Psi_{iP}-\frac{3}{2}C_{MN}C^{MN}
-\frac{3}{2}k\epsilon(y)\bar{\Psi}^i_M\gamma^{MN}\Psi_{iN}\,
-12k^2+...\right) \right.\nonumber \\
&&\quad+\,\,\frac{1}{{g}_{5a}^2}
\left(-\frac{1}{4}F^{aMN}F^a_{MN}
+\frac{1}{2}D_M\phi^aD^M\phi^a
+\frac{i}{2}\bar{\lambda}^{ai}\gamma^MD_M\lambda^a_i
+\frac{1}{2}k\epsilon(y)\bar{\lambda}^{ia}\lambda^a_i-
4k^2\phi^a\phi^a
+...\right)\nonumber \\
&&\left.\quad\quad\quad+\,\,
\Big(\,|D_Mh_I^i|^2+i\bar{\Psi}_I\gamma^MD_M\Psi_I+
iM_I\epsilon(y)\bar{\Psi}_I\Psi_I+
(M_I^2\pm kM_I-\frac{15}{4}k^2)|h^i_I|^2+...\,\Big)\,\right]\,,
\eea
where ${\cal R}$ is the Ricci scalar of the 5D metric
$G_{MN}$, $\Psi^i_M$ ($i=1,2$) are
$SU(2)_R$-doublet 5D gravitino,
$C_{MN}=\partial_MB_N-\partial_NB_M$ is the graviphoton field strength.
The fields $(\phi^a,A_M^a,\lambda^{ia})$ are 5D real scalar, vector and
$SU(2)_R$-doublet gaugino constituting a vector multiplet,
$(h_I^i,\Psi_I)$ are $SU(2)_R$-doublet complex hyperscalar and
Dirac hyperino constituting
a  hypermultiplet, and
$\epsilon(y)=\pm 1$ is the periodic sign function on $S^1/Z_2$
satisfying $\epsilon(y)=\epsilon(y+2\pi)=-\epsilon(-y)$.
Here we set the 5D Planck mass $M_5=1$, and the ellipses
include appropriate boundary actions.

In 5D orbifold SUGRA, the bulk and boundary cosmological constants
%$-6k^2$
and also the hyperino
kink masses
%$M_I\epsilon(y)$
appear in connection with the
graviphoton gauge couplings
\cite{Choi:2002wx}:
\bea
\label{gauging}
&& D_M h^i_I=\nabla_Mh^i_I-
i\Big(\frac{3}{2}k(\sigma_3)^i_j-
c_I\delta^i_j\Big)\epsilon(y)B_Mh^j_I,
\nonumber \\
&& D_M\Psi_I=\nabla_M\Psi_I+ic_I\epsilon(y)B_M\Psi_I,
\nonumber \\
&& D_M\lambda^{ia}=\nabla_M\lambda^{ia}-i\frac{3}{2}k(\sigma_3)^i_j
\epsilon(y)B_M\lambda^{aj},
\eea
where $\nabla_M$ contains other gauge couplings.
For instance, the $Z_2$-odd $U(1)_R$ gauge coupling of
the graviphoton, $\frac{3}{2}k\epsilon(y)\sigma_3$, is associated with
the bulk and boundary cosmological constants:
$$
-6k^2+6k\frac{(\delta(y)-\delta(y-\pi))}{\sqrt{G_{55}}}
$$
which leads to the warped Randall-Sundrum geometry.
As for the hyperino kink mass $M_I\epsilon(y)$, another possible origin
is a $U(1)_{FI}$ vector multiplet whose scalar component
develops a kink-type vacuum expectation value
due to the boundary FI terms \cite{Barbieri:2002ic}.
Including this FI contribution, the (effective) hyperino kink mass is given by
\bea
M_I=c_I+q_I\xi_{FI}\,,
\eea
where $q_I$ is the $U(1)_{FI}$ charge of $\Psi_I$ and
$\xi_{FI}$ is the FI coefficient.
Throughout this paper, we will assume that the $U(1)$ gauge charges
$c_I,q_I$ are {\it quantized},
and thus the hyperino kink masses are quantized also.

In 5 dimension,
%space transformations for orbifolding are given by
%\bea
%{\cal Z} : y \rightarrow -y\,,
%\nonumber \\
%{\cal T}: y \rightarrow y+2\pi.
%\eea
orbifolding the theory corresponds to imposing the boundary condition
\bea
\Phi(-y)=Z\Phi(y)\,,\quad
\Phi(y+2\pi)=\Omega\Phi(y)\,,
\label{bc1}
\eea
for generic 5D field $\Phi$, where $Z$ and $\Omega$
%are the representation of ${\cal Z}$ and ${\cal T}$, respectively,
%in the space of generic 5D fields $\Phi$. To be consistent, one needs
satisfy the consistency conditions
\beq
Z^2=1\,,\quad
Z\Omega=\Omega^{-1}Z.
\label{consistency1}
\eeq
Equivalently, one can impose the parity boundary condition
at each fixed point:
\bea
\Phi(-y)=Z\Phi(y)\,,\quad
\Phi(-y')=Z'\Phi(y') \quad
(Z^2=Z^{\prime 2}=1)\,,
\label{bc2}
\eea
where $y'=y-\pi$ and $Z'=Z\Omega$.
In case that
$Z$ and $\Omega$ commute to each other, the
consistency condition (\ref{consistency1}) implies
$\Omega^2=1$, and thus
$Z$ and $\Omega$ can be simultaneously diagonalized
to have eigenvalues $Z=\pm 1$ and $\Omega=\pm 1$.
On the other hand, for the case of SS boundary condition,
$Z$ and $\Omega$ do not commute,
so $\Omega$ can have a continuous value.
Still one can adopt a field basis for which
$Z$ is diagonal, while $\Omega$ (and thus $Z'=Z\Omega$)
is not diagonal in general.

The orbifolding boundary condition
(\ref{bc1}) should be consistent with
all gauge symmetries of the theory, including
the $Z_2$-odd graviphoton gauge transformation:
\bea
B_M\,\rightarrow \,B_M+\partial_M\Lambda,
\quad
\Phi\,\rightarrow \,e^{i\epsilon(y)Q\Lambda}\Phi,
\eea
where $Q$ is a constant charge matrix
and the transformation function satisfy
$\Lambda(y)=\Lambda(y+2\pi)=-\Lambda(-y)$
in order to be consistent with 5D local SUSY.
In order for $D_M\Phi=(\nabla_M-i\epsilon(y)QB_M)\Phi$
to have a consistent boundary condition,
both $Z$ and $\Omega$ should commute with $Q$:
\bea
ZQ=QZ,
\quad
\Omega Q=Q\Omega
\label{consistency2}
\eea
which correspond to additional consistency condition
for orbifolding boundary conditions.

5D orbifold SUGRA admits a continuous twist of $SU(2)_R$ boundary condition
under the discrete shift $y\rightarrow y+2\pi$,
breaking the $N=1$ SUSY survived from the $Z_2$-orbifolding.
Let $Z_R,\Omega_R$ and $Q_R$ denote the $SU(2)_R$ representations of
$Z,\Omega$ and $Q$, respectively.
To obtain 4D chiral fermion, one can always choose $Z_R=\sigma_3$,
and then the graviphoton $U(1)_R$ charge
is given by $Q_R=\frac{3}{2}k\sigma_3$ as in
(\ref{gauging}).
The consistency condition (\ref{consistency1})
implies that a continuous SS twist can be written as
$\Omega_R=\exp(i\vec{\omega}\cdot\vec{\sigma})$
where $\vec{\omega}=(\omega_1,\omega_2,0)$.
However, if $Q_R=\frac{3}{2}k\sigma_3$
is non-vanishing,
there doesn't exist any non-trivial SS twist
allowed by
the consistency condition
(\ref{consistency2}).
In other words, a continuous SS SUSY breaking
is not allowed in 5D orbifold SUGRA yielding a
warped Randall-Sundrum geometry as has been noticed
in \cite{Hall:2003yc}.
Thus in the following, we will focus on the case that
the graviphoton $U(1)_R$ charge vanishes:
\bea
Q_R=\frac{3}{2}k\sigma_3=0,
\nonumber
\eea
which gives a flat spacetime geometry
\bea
ds^2=\eta_{\mu\nu}dx^\mu dx^\nu-R^2dy^2
\nonumber
\eea
and allows a continuous SS twist
\bea
\Omega_R=\exp (2\pi i\omega\sigma_2)
\nonumber
\eea
in an appropriate $SU(2)_R$
basis.
Obviously, only the $SU(2)_R$-doublet gravitino, gauginos
and hyperscalars are affected by this SS twist, e.g.
\bea
&&\lambda^{ai}(y+2\pi)=\left(e^{2\pi i{\omega}\sigma_2}
\right)^i_j
\lambda^{aj}(y)\,,\nonumber\\
&& h^i_I(y+2\pi)=\left(e^{2\pi i{\omega}\sigma_2}\right)^i_j
h^j_I(y)\,.
\label{sstwist}
\eea

It is convenient to write the  5D action (\ref{5daction1}) in $N=1$
superspace \cite{Arkani-Hamed:2001tb}.
For the 5D SUGRA multiplet, we keep only the radion superfield
$$
T=R+iB_5+\theta \Psi_{5R}+
\theta^2F^T\,,
$$
where $R=\sqrt{G_{55}}$ denotes the radius of the compactified
5-th dimension, and
$\Psi_{5R}=
\frac{1}{2}(1+\gamma_5)\Psi^{i=2}_{M=5}$.
The relevant piece of the 5D action in $N=1$ superspace is given by
\bea
\int d^5x \,\left[\,
\int d^4 \theta \,\frac{T+T^*}{2}\left( H_I H^*_I+H^c_I H^{c*}_I\right)
+ \left(\,\int d^2\theta\,
H^c_I\left(\partial_y+M_I T\epsilon(y)\right)H_I
+\frac{1}{4g^2_{5a}}T W^{a\alpha} W^a_{\alpha} + {\rm h.c.}\right)
\,\right],
\label{bulkaction}
\eea
where $W^{a}_{\alpha}$ is the
chiral spinor superfield for the 5D vector superfield
$$
{V}^a=-\bar{\theta}\sigma^\mu\theta A^a_\mu
-i\bar{\theta}^2\theta\lambda^a+i\theta^2\bar{\theta}\bar{\lambda}^a
+\frac{1}{2}\theta^2\bar{\theta}^2D^a\,,
$$
and
\bea
\label{hyper}
H_I&=&h^{1}_I+\theta \psi_I+\theta^2 F^{1}_I\,,
\nonumber \\
H_I^{c}&=&h^{2*}_I+
\theta\psi^c_I+\theta^2 F^{2*}_I\,,
\eea
for
$\lambda^a=\frac{1}{2}(1-\gamma_5)\lambda^{a1}$,
$\psi_I=\frac{1}{2}(1-\gamma_5)\Psi_I$,
and $\bar{\psi^c}_I=\frac{1}{2}(1+\gamma_5)\Psi_I$.

Here we are interested in 5D vector multiplets giving
massless 4D gauge bosons, and also 5D hypermultiplets
giving massless chiral 4D fermions.
We thus consider the $N=1$ superfields satisfying
the $Z_2$-boundary
condition:
\beq
{V}^a(-y)={V}^a(y),\quad
H_I (-y) =H_I (y),\quad
H_I^c(-y) = -H^c_I (y),
\label{superbc}
\eeq
The zero mode equation for $\psi_I=
\chi(x)\tilde{\phi}_{0I}(y)$ is given by
\bea
\big(\,\partial_y+M_IT\epsilon(y)\,\big)\tilde{\phi}_{0I}=0,
\nonumber
\eea
yielding the zero mode wavefunction
$$
\tilde{\phi}_{0I}\propto e^{-M_IT|y|},
$$
which shows that the  zero mode is quasi-localized at $y=0$ if $M_I>0$,
and at $y=\pi$ if $M_I<0$.

In 5D orbifold SUGRA, Yukawa couplings can be introduced
only through the boundary actions.
For the orbifolding given by (\ref{bc2}),
the boundary action at $y=0$ is
required to be invariant under the $Z$-even
supercharges ${\cal Q}_Z$, while the boundary action at $y=\pi$
is invariant under the $Z'$-even supercharges ${\cal Q}_{Z'}$.
Note that ${\cal Q}_Z$
and ${\cal Q}_{Z'}$ are related to each other
by the SS twist: ${\cal Q}_Z=\Omega_R^{1/2}{\cal Q}_{Z'}$.
Then  in the presence of nontrivial SS twist,
the boundary actions for Yukawa couplings can be written as
\beq
\int d^5 x \int d^2 \theta \,\Big(
\,\delta(y) \frac{1}{6}\lambda_{IJK} H_I H_J H_K +
\delta(y-\pi) \frac{1}{6}
\lambda^{\prime}_{IJK} H^\prime_I H^\prime_J H^\prime_K\,
\Big) + {\rm h.c}\,,
\label{boundaryaction}
\eeq
where $H_I$ and $H^c_I$ are $Z$-even and $Z$-odd
superfields, respectively, defined as (\ref{hyper})
for the $N=1$ SUSY generated by ${\cal Q}_Z$,
and $H'_I$ and $H^{\prime c}_I$
are $Z'$-even and $Z'$-odd superfields, respectively,
for the $N=1$ SUSY generated by ${\cal Q}_{Z'}$.
More explicitly,
\bea
H'_I &=& h^{\prime 1}_I+\theta\psi_I+\theta^2 F^{\prime 1}_I, \quad
\nonumber \\
H^{\prime c}_I &=& h^{\prime 2*}_I+\theta\psi^c_I+\theta^2F^{\prime 2*}_I,
\nonumber
\eea
where
\bea
h^{\prime 1}_I &=& \cos(\omega\pi) h^1_I - \sin(\omega\pi) h^{2}_I,
\nonumber \\
h^{c\prime}_I &=& \cos(\omega\pi) h^{2*}_I+\sin(\omega\pi) h^{1*}_I,
\nonumber
%F^{\prime 1}_I &=& \cos(\omega\pi) F^1_I - \sin(\omega\pi) F^{2}_I,
%\nonumber \\
%F^{c\prime}_I &=& \sin(\omega\pi) F^{2*}_I + \cos(\omega\pi) F^{1*}_I
\eea
and $F^{\prime i}_I$ can be determined by their equations of motion.
Note that $Z$-odd $H^c_I$ and $Z'$-odd
$H^{\prime c}_I$
vanish at $y=0$ and $y=\pi$, respectively,
thus the boundary Yukawa operators involving $H^c_I$ ($H^{\prime c}_I$)
at $y=0$ ($y=\pi$) vanish also.

Let us now compute the 4D Yukawa couplings,
scalar masses and trilinear
$A$ parameters for the hypermultiplets
obeying  the boundary conditions
(\ref{sstwist}) and (\ref{superbc}).
To this end, we analyze the Kaluza-Klein (KK) mass spectrum and
wavefunctions of hypermultiplets, which has been done in \cite{quiros}.
The equation of motion for hyperscalar fields
leads to the following KK wave equation:
\beq
\label{eom}
\left(
\frac{1}{R^2}\partial^2_y+m^2_I -M_I^2+ \frac{M_I}{R}
\sigma_3\partial_y\epsilon(y)
\right)\tilde{\phi}_I(y)
=0,
\eeq
where $\partial_y\epsilon(y)=
2(\delta(y)-\delta(y-\pi))$, and
the KK wavefunction $\tilde{\phi}_I=(\tilde{\phi}^1_I,\tilde{\phi}^2_I)$
is defined as
\bea
h^i_I(x,y)=\phi_I(x)\tilde{\phi}^i_I(y)
\nonumber
\eea
for the 4D field $\phi_I(x)$ satisfying the on-shell condition
$$
\partial_\mu\partial^\mu\phi_I(x)=m_I^2\phi_I(x).
$$
According to the boundary conditions (\ref{sstwist})
and (\ref{superbc}),  $\tilde{\phi}_I$ obeys
\bea
\label{sstwist1}
\tilde\phi^i_I(-y)=\big(\sigma_3\big)^i_j\tilde{\phi}^j_I(y)\,,
\quad
\tilde{\phi}^i_I(y+2\pi) = \big(
e^{i2\pi\omega\sigma_2}\big)^i_j
\tilde{\phi}^j_I(y).
\eea
It is then straightforward to find \cite{quiros}
\bea
\tilde{\phi}^1_{I} &=&
C_I \left(\cos {\Delta_I}y - \frac{M_I R}{\Delta_I} \sin \Delta_I y \right),
\nonumber \\
\tilde{\phi}^2_{I} &=&
C_I \tan \omega \pi \left(-\cot \Delta_I\pi + \frac{M_I R}{\Delta_I} \right)
\sin \Delta_I y,
\label{kkwavefunction}
\eea
where $\Delta_I$ satisfies
\beq
\left(\frac{R M_I}{\Delta_I}\right)^2+1 =
\frac{\sin^2 \omega \pi}{\sin^2 \Delta_I \pi}\,.
\nonumber
\eeq
The corresponding KK mass eigenvalue
is given by
\bea
\label{kkmass}
m_I^2=M_I^2+\left(\frac{\Delta_I}{R}\right)^2\,,
\eea
and $C_I$ is the normalization constant
which can be determined by
$$2R\int_0^\pi \,dy \,\,\big(\,
|\tilde{\phi}^1_{I}|^2+|\tilde{\phi}^2_{I}|^2\,\big)=1\,.
$$
Note that (\ref{kkwavefunction}) represents
the KK wavefunction over the fundamental domain $0\leq y\leq \pi$.
The KK wavefunction outside the fundamental domain can be determined by
the boundary condition (\ref{sstwist1}).

Once the KK wavefunctions are determined,
the 4D couplings of the corresponding KK modes can be easily obtained.
Let $y_{IJK}$ and $A_{IJK}$
denote the 4D Yukawa couplings and trilinear scalar couplings,
respectively,
of the {\it canonically normalized}
4D scalars $\phi_I$ and 4D  fermions $\chi_I$:
\bea
\int d^4x \,\Big(\,\frac{1}{2}y_{IJK}\phi_I\chi_J\chi_K
-\frac{1}{6}A_{IJK}\phi_I\phi_J\phi_K+{\rm h.c.}
\Big).
\eea
Since the hyperinos are not affected by the SS twist,
the hyperino KK wavefunctions correspond to
the hyperscalar KK wavefunctions with $\omega=0$, i.e.
\bea
h^i_I(x,y)=\phi_I(x)\tilde{\phi}_I^i(y),
\quad
\psi_I(x,y)=\chi_I(x)\tilde{\phi}_{0I}^i(y),
\nonumber
\eea
where $\tilde{\phi}_{0I}^i\equiv \tilde{\phi}^i_I|_{\omega=0}$.
In our case, the 4D Yukawa couplings of KK modes are easily found
to be
\bea
\label{4dyukawa}
y_{IJK} &=&
\int \,dy \,\Big( \,
\lambda_{IJK} \delta(y)\tilde{\phi}^1_{I} \tilde{\phi}^1_{0J}
\tilde{\phi}^1_{0K} +
\lambda^\prime_{IJK}\delta(y-\pi)
\tilde{\phi}^{\prime 1}_{I}\tilde{\phi}^1_{0J}
\tilde{\phi}^1_{0K} \,
\Big)\,,
\eea
where
\bea
\tilde{\phi}^{\prime i}_I =
\big(e^{i\pi\omega\sigma_2}\big)^i_j\tilde{\phi}^j_I\,.
\nonumber
\eea
Using the equation of motion for the auxiliary components
$F^i_I$:
\begin{eqnarray}
F^i_I&=&\frac{1}{R}\big(i\sigma_2\big)^i_j\partial_yh_I^j
-M_I\epsilon(y)\big(\sigma_1\big)^i_jh^j_I\,,
\nonumber \\
F^{\prime i}_I&=&\frac{1}{R}\big(i\sigma_2\big)^i_j\partial_yh_I^{\prime j}
-M_I\epsilon(y)\big(\sigma_1\big)^i_jh^{\prime j}_I\,,
\nonumber
\end{eqnarray}
one can obtain also the trilinear soft scalar couplings
as
\bea
\label{4da}
A_{IJK} &=&
-\frac{1}{R}\int\, dy \,
\Big(\,\lambda_{IJK} \delta(y)\tilde{\phi}^1_{I}
\tilde{\phi}^1_J
\partial_y\tilde{\phi}^2_{K} +
\lambda^\prime_{IJK}\delta(y-\pi)
\tilde{\phi}^{\prime 1}_{I}
\tilde{\phi}^{\prime 1}_{J}\partial_y\tilde{\phi}^{\prime 2}_{K}
\Big)
\nonumber \\
&&\quad\quad\quad\quad\quad + \,\,
\Big(\,I\leftrightarrow K\,\Big) +
\Big(\,J\leftrightarrow K\,\Big)\,.
\eea

The scalar masses and trilinear couplings discussed above can be interpreted as
the parameters renormalized at the compactification scale $M_c\sim 1/R$.
It is rather clear that the resulting soft parameters of quasi-localized matter zero
modes can not be phenomenologically viable unless $M_c$ is far above the weak scale
$M_W$.
Since the soft parameters have values of ${\cal O}(\omega M_c)$,
one needs $|\omega|\ll 1$ to get the weak scale soft parameters
for $M_c\gg M_W$.
With this observation, in the following, we limit the discussion to the case of
{\it small} SS parameter,
\bea
|\omega|\,\ll\, 1\,,
\nonumber
\eea
and compute the soft parameters of
the canonically normalized
hyperscalar zero modes $\phi_I$ and the gaugino zero modes
$\lambda^a$:
\bea
-\int d^4x \, \Big(\,
\frac{1}{2}m^2_{IJ}\phi^*_I\phi_J+\frac{1}{6}
A_{IJK}\phi_I\phi_J\phi_K+
\frac{1}{2}M_a\lambda^a\lambda^a+{\rm h.c.}\,\Big).
\eea
Under the SS twist (\ref{sstwist}), the gaugino zero mode
receives a soft mass
\bea
M_a=-\frac{\omega}{R}.
\nonumber
\eea
From (\ref{kkwavefunction}) and (\ref{kkmass}), one easily finds
the KK wavefunctions of the hyperscalar zero modes:
\bea
\tilde{\phi}^1_{I} &=&
\left(\frac{M_I}{1-e^{-2M_I\pi R}}\right)^{1/2}e^{-M_IRy}+{\cal O}
(\omega^2),
\nonumber \\
\tilde{\phi}^2_{I} &=& \pi\omega\left(\frac{M_I}{
e^{2M_I\pi R}-1}\right)^{1/2}
\frac{e^{M_I R y}-e^{-M_I R y}}{e^{M_I\pi R}-
e^{-M_I\pi R}}+{\cal O}(\omega^3),
\eea
and their soft masses:
\beq
m^2_{IJ}=m^2_I\delta_{IJ}=
\left( \frac{\omega}{R}\right)^2\left( \frac{M_I \pi R}{\sinh (M_I \pi R)}\right)^2
+{\cal O}(\omega^4).
\eeq
The Yukawa couplings and A-parameters of these zero modes
can be obtained from (\ref{4dyukawa}) and (\ref{4da}), yielding
\bea
y_{IJK} &=&
\frac{1}{\sqrt{Y_IY_JY_K}}
\Big(\,
\lambda_{IJK}+\lambda^\prime_{IJK}e^{-(M_I+M_J+M_K)\pi R}\,
\Big)+{\cal O}(\omega^2),
\nonumber \\
A_{IJK} &=&
\frac{\omega}{R}\frac{1}{\sqrt{Y_IY_JY_K}}
\left[\left(
\frac{2M_I \pi R}{e^{2M_I\pi R}-1}+\frac{2M_J \pi R}{e^{2M_J\pi R}-1}+
\frac{2M_K \pi R}{e^{2M_K\pi R}-1}
\right)\lambda_{IJK}
\right.
\nonumber \\
&+&\left. \left(
\frac{2M_I \pi R}{1-e^{-2M_I \pi R}}+\frac{2M_J \pi R}{1-e^{-2M_J \pi R}}+
\frac{2M_K \pi R}{1-e^{-2M_K \pi R}}
\right)\lambda^\prime_{IJK}e^{-(M_I+M_J+M_K)\pi R} \right]
+{\cal O}(\omega^3)\,,
\eea
where
$$
Y_I=\frac{1}{M_I}\big(\, 1- e^{-2M_I\pi R}\,\big).
$$
Note that the above Yukawa couplings and soft parameters of zero modes
should be interpreted as the parameters renormalized at the
compactification scale $M_c\sim 1/R$.

It has been pointed out that the SS SUSY breaking in 5D orbifold SUGRA
has an interesting correspondence with the radion-mediated SUSY breaking.
Here we explicitly show that the zero mode soft parameters
induced by the SS boundary condition
are precisely same as the radion-mediated
soft parameters in 4D effective theory.
To see this, let us construct the
4D effective action of the gauge and matter zero modes
without any SS twist, i.e. $\omega=0$, while keeping
the radion superfield $T$ to take a generic value.
The resulting 4D effective action can be written on $N=1$ superspace,
and can be obtained easily
by making the radion-dependent superfield redefinition:
\bea
{H}_I \,\rightarrow\, e^{M_I T |y|} H_I, \quad
{H}^c_I \,\rightarrow\, e^{-M_I T |y|} H^c_I. \nonumber
\eea
After this field redefinition,
the bulk and boundary actions of (\ref{bulkaction}) and
(\ref{boundaryaction})
become
\bea
\label{5d action}
{S}_{\mathrm{bulk}} &=&
\int d^5 x  \left[\, \int d^4 \theta \, \frac{T+T^*}{2} \left(
e^{-M_I(T+T^*)|y|} {H}_I {H}^*_I +
e^{M_I(T+T^*)|y|} {H}^c_I {H}^{c*}_I \right) \right.
\nonumber \\
&&
\left. \hspace{10mm}
+  \int d^2 \theta
\frac{1}{4g^2_{5a}} T W^{a\alpha} W^a_{\alpha}+ {\rm h.c.} \,
\right],
\nonumber \\
{S}_{\mathrm{brane}} &=&
\int d^5 x \int d^2 \theta
\left( \,\delta(y) \frac{1}{6}\lambda_{IJK} {H}_I {H}_J {H}_K
+ \delta(y-\pi)\frac{1}{6} \lambda^{\prime}_{IJK}
e^{-(M_I+M_J+M_K)T|y|}{H}_I {H}_J {H}_K\, \right)
+ {\rm h.c.}
\eea
In the new 5D superfield basis, all zero modes
have {\it constant} wavefunctions, thus
their 4D effective action can be obtained by simply
integrating the 5D action over the 5-th dimension.
Let $\Phi_I$ denote the constant zero modes of $H_I$, and
$W^a_\alpha$ denote the field strength superfields for the
constant zero modes of
$V^a$. We then find
\bea
\label{4d action}
{S}_{4D} =
\int d^4 x \left[
\int d^4 \theta \, Y_{I\bar{J}} \Phi_I \Phi^*_J
+ \int d^2 \theta \left(
\frac{1}{4} f_a W^{a\alpha} W^a_{\alpha} + \tilde{y}_{IJK} \Phi_I
\Phi_J \Phi_K
\right) + {\rm h.c.}\,  \right],
\eea
where   the hermitian wavefunction coefficients $Y_{I\bar{J}}$,
the holomorphic Yukawa couplings
$\tilde{y}_{IJK}$, and
the holomorphic
gauge kinetic functions $f_a$ are given by
\bea
Y_{I\bar{J}} &=&
Y_I \delta_{IJ} =
\frac{1}{M_I} \left( 1 - e^{-\pi M_I (T+T^*)} \right)
\delta_{IJ},
\nonumber \\
\tilde{y}_{IJK} &=&
\lambda_{IJK} + \lambda^{\prime}_{IJK} e^{-\pi(M_I+M_J+M_K)T},
\nonumber \\
f_a &=& \frac{2\pi}{g^2_{5a}}T.
\eea
If the radion $F$-component, $F^T$,
is the major source of SUSY breaking in the above 4D effective action,
the soft parameters of the canonically normalized
4D fields are given by
\bea
M_a &=& -\frac{1}{2 \, \mathrm{Re}(f_a)} \, F^T \partial_T f_a
\,\,=\,\,-\frac{F^T}{2R}\,,
\nonumber \\
m^2_{I\bar{J}} &=&
- \frac{1}{\sqrt{Y_I Y_J}} \left| F^T \right|^2
\left( \partial_T \partial_{T^*} Y_{I \bar{J}}
-{\Gamma_T}^K_I \partial_{T^*} Y_{K \bar{J}} \right)
\nonumber \\
&=&\delta_{IJ}\left| \frac{F^T}{2R}\right|^2
\left( \frac{M_I \pi R}{\sinh (M_I \pi R)}\right)^2,
\nonumber \\
A_{IJK} &=& - \frac{1}{\sqrt{Y_I Y_J Y_K}} \, F^T
\left(
\partial_T {\tilde{y}}_{IJK}
- {\Gamma_T}^L_I {\tilde{y}}_{LJK}
- {\Gamma_T}^L_J {\tilde{y}}_{ILK}
- {\Gamma_T}^L_K {\tilde{y}}_{IJL}
\right)
\nonumber \\
&=&\frac{F^T}{2R}\frac{1}{\sqrt{Y_IY_JY_K}}
\left[\left(
\frac{2M_I \pi R}{e^{2M_I\pi R}-1}+\frac{2M_J \pi R}{e^{2M_J\pi R}-1}+
\frac{2M_K \pi R}{e^{2M_K\pi R}-1}
\right)\lambda_{IJK}
\right.
\nonumber \\
&+&\left. \left(
\frac{2M_I \pi R}{1-e^{-2M_I \pi R}}+\frac{2M_J \pi R}{1-e^{-2M_J \pi R}}+
\frac{2M_K \pi R}{1-e^{-2M_K \pi R}}
\right)\lambda^\prime_{IJK}e^{-(M_I+M_J+M_K)\pi R} \right],
\eea
where the K$\ddot\mathrm{a}$hler connection
${\Gamma_T}^I_J = Y^{I\bar{K}} \partial_T Y_{J\bar{K}}$.
Obviously
the above radion-mediated soft
parameters are precisely same as
the SS-induced soft parameters (up to small
corrections higher order in  $\omega$)
with the matching condition
$$
F^T=2\omega\,.
$$
This means that our phenomenological analysis of
SS-induced soft parameters in the next section
can be applied  to any SUSY breaking mechanism
giving a sizable $F^T$,
for instance the hidden gaugino condensation
model.

So far, we have considered the most general scenario that
the Yukawa couplings originate from both fixed points, $y=0$ and $\pi$.
In fact, to generate hierarchical Yukawa couplings through
quasi-localization in a natural manner,
one needs to assume that Yukawa couplings
originate {\it only} from one fixed point, e.g. from $y=0$.
In this case, the Yukawa couplings  and soft parameters
at the compactification scale are given by
\bea
\label{result}
y_{IJK} &=&
\left(
\frac{M_I M_J M_K}{(1-e^{-2M_I \pi R})(1-e^{-2M_J \pi R})
(1-e^{-2M_K \pi R})
}\right)^\frac{1}{2} \lambda_{IJK}
\nonumber \\
A_{IJK} &=&
\frac{\omega}{R}
\left(
\frac{2M_I \pi R}{e^{2M_I \pi R}-1}+\frac{2M_J \pi R}{e^{2M_I \pi R}-1}
+\frac{2M_K \pi R}{e^{2M_K \pi R}-1}
\right) y_{IJK}
\nonumber \\
M_a &=& -\frac{\omega}{R}\,,
\nonumber \\
m_I^2&=&
\left( \frac{\omega}{R}\right)^2
\left( \frac{M_I \pi R}{\sinh (M_I \pi R)}\right)^2 .
\eea
As we have noticed, the matter zero mode $\Phi_I$
from a 5D hypermultiplet with kink mass $M_I$
has an wavefunction of the form $e^{-M_IR|y|}$,
thus is quasi-localized at $y=0$ ($y=\pi$)
if $M_I>0$ ($M_I<0$).
As a result, in case that Yukawa couplings
originate from  $y=0$,
the quark/lepton superfields from hypermultiplets with
$M_I<0$ would have
(exponentially) small Yukawa couplings, while the
quark/lepton superfields
with $M_I>0$ can have Yukawa couplings of order unity.
Obviously, the above form of Yukawa couplings shows
this feature, achieving
the Yukawa hierarchy from quasi-localization.
The above results  show also that
the soft scalar masses $m_I^2$
and the $A$ to Yukawa ratios $A_{IJK}/y_{IJK}$
are highly flavor-dependent at the compactification scale.
Although the flavor-violating pieces are suppressed
with an appropriate correlation with Yukawa couplings,
still they can lead to dangerous flavor-violations
at low energy scales as will be discussed in the next section.

Assuming that the Higgs superfields
are boundary superfields confined
at $y=0$ simplifies the form of Yukawa and $A$-parameters, however
their flavor structures are essentially the same.
In our framework, any boundary superfield can be interpreted
as the zero mode of bulk hypermultiplet having an infinite kink mass,
more precisely a kink mass comparable to
the 5D cutoff scale $\Lambda_{5}$.
Note that in this case all other KK modes have
the masses of ${\cal O}(\Lambda_{5})$, so are decoupled.
Then the Yukawa and $A$-parameters of boundary Higgs superfields
$\Phi_K$ can be  obtained from (\ref{result}) by taking the limit
$M_K\rightarrow\Lambda_5$ together with an appropriate redefinition
of boundary Yukawa couplings, yielding
\bea
y_{IJ} &=&
\left[
\frac{M_IM_J}{(1-e^{-2M_I \pi R})(1-e^{-2M_J \pi R})
}\right]^\frac{1}{2} \tilde{\lambda}_{IJ}
\nonumber \\
A_{IJ} &=&
\frac{\omega}{R}
\left(
\frac{2M_I \pi R}{e^{2M_I \pi R}-1}+\frac{2M_J \pi R}{e^{2M_I \pi R}-1}
\right) y_{IJ}\,
\eea
where
\bea
\tilde{\lambda}_{IJ}\,\,\equiv\,\,
\left(\frac{M_K}{1-e^{-2M_K\pi R}}\right)^{1/2}\lambda_{IJK}
\,\,\approx\,\, \Lambda_5^{1/2}\lambda_{IJK}\,
\label{boundaryhiggs}
\eea
for the quark/lepton flavor indices $I,J$.

In the next section, we analyze in detail the
resulting low energy flavor violations under the assumption
that the kink masses  $M_I$ are appropriately quantized.
Note that in 5D orbifold SUGRA
the assumption of quantized kink masses
corresponds to the assumption of quantized $U(1)$ gauge charges.
It is then convenient to write
the above Yukawa couplings and soft parameters
in the following way:
\bea
\label{radiondomination}
y_{IJ}&=&\tilde\lambda_{IJ}\frac{\ln(1/\epsilon)}{\pi R}
\sqrt{\frac{N_IN_J}{(\epsilon^{-2N_I}-1)(
\epsilon^{-2N_J}-1)}}\,,
\nonumber \\
M_a&=& -\frac{\omega}{R}\,,\nonumber \\
A_{IJ}&=&2y_{IJ}\ln(1/\epsilon)\frac{\omega}{R}\left(\,
\frac{N_I}{1-\epsilon^{2N_I}}+
\frac{N_J}{1-\epsilon^{2N_J}}\,\right),
\nonumber \\
m^2_{I\bar{J}}&=&\delta_{IJ}\left(\,
2\ln(1/\epsilon)\frac{N_I}{\epsilon^{N_I}-
\epsilon^{-N_I}}\frac{\omega}{R}\,\right)^2,
\eea
where
$$
N_I=-\frac{\pi R}{
\ln(1/\epsilon)}M_I\,
\quad\mbox{for}\quad \epsilon\,\equiv \,\mbox{Cabibbo angle}\,
\approx 0.2.
$$

The above Yukawa couplings are quite similar to
the Yukawa couplings in Frogatt-Nielsen models
with $N_I$ being identified as the $U(1)_F$ charges.
More explicitly,
\bea
y_{IJ}\,\simeq \,
\frac{\tilde{\lambda}_{IJ}\ln(1/\epsilon)}{\pi R}\,
\sqrt{Z_IZ_J}\,\epsilon^{X_I+X_J}\,\equiv\,
\lambda_{IJ}\epsilon^{X_I+X_J}\,,
\label{lambda}
\eea
where
\beq
Z_I=
\left\{
\begin{array}{ll}
|N_I| & \quad (\,N_I \neq 0\,) \\
1/[2\ln(1/\epsilon)] & \quad (\,N_I =0\,)\,,
\end{array}
\right.
\nonumber
\eeq
and the effective flavor charge $X_I$ is given by
\beq
X_I =
\left\{
\begin{array}{ll}
N_I & \quad (\,N_I \geq 0\,) \\
0 & \quad (\,N_I <  0\,)\,.
\end{array}
\right.
\nonumber
\eeq
When the theory is strongly
coupled at $\Lambda_5$,
a naive dimensional analysis \cite{luty} suggests that
$$
\Lambda_5\pi R={\cal O}(6\pi ^3),
\quad
\tilde{\lambda}_{IJ}={\cal O}(\sqrt{6\pi ^3}/\Lambda_5).$$
Then the redefined boundary Yukawa couplings
$\lambda_{IJ}\simeq
\tilde{\lambda}_{IJ}\sqrt{M_IM_J}$
would be of order unity if the corresponding kink masses
$|M_I|=|N_I|\ln(1/\epsilon)/\pi R={\cal O}(\sqrt{6\pi^3}/\pi R)$.
In the following, we will ignore the factor $2\sim 3$ differences of $\lambda_{IJ}$
arising from their $M_I$-dependence, and simply assume that the redefined boundary Yukawa couplngs
$\lambda_{IJ}$ are all of order unity.
Then the observed quark/lepton masses and CKM mixing angles
can be explained by the 5D kink masses $M_I$ which are quantized
in a manner to give integer-valued $N_I$.

As for the soft scalar masses and trilinear couplings, the
above results
can be approximated as
\begin{eqnarray}
A_{IJ}&\,\simeq\, & M_0\, y_{IJ}\,
(a_I+a_J)\,,
\nonumber \\
m^2_{I\bar{J}}&\simeq &\delta_{I\bar{J}} M_0^2
\left\{
\begin{array}{ll}
N_I^2\epsilon^{2|N_I|} &~~~(\,N_I \neq 0\,)\\
1/[2\ln(1/\epsilon)]^2 &~~~(\,N_I = 0\,)
\label{soft1}
\end{array}
\right.
\end{eqnarray}
where
\beq
M_0=2\ln(1/\epsilon)\frac{\omega}{R},
\nonumber
\eeq
and
\beq
a_I =
\left\{
\begin{array}{ll}
N_I & \quad (\,N_I > 0\,) \\
1/[2\ln(1/\epsilon)] & \quad (\,N_I = 0\,)
\\
|N_I|\epsilon^{2|N_I|} &
\quad (\, N_I<0\,)\,.
\end{array}
\right.
\nonumber
\eeq
An important feature of the SS SUSY breaking
is that $A_{IJ}/y_{IJ}$ and $m^2_{I\bar{J}}$ are {\it quantized}
for $N_I\neq 0$ in the leading approximation. This feature of
the SS SUSY breaking, more generally of the radion-mediated SUSY breaking,
is quite useful for suppressing dangerous flavor violations.
With this feature, flavor violating amplitudes appear in a form
$f(N_I)-f(N_J)$, thus are canceled
if some of $N_I$ are degenerate.

The suppression
of $m^2_{I\bar{J}}/M_0^2$ and $A_{IJ}/M_0$  by some powers of $\epsilon$
is essentially due to the quasi-localization of
matter zero modes.
The SUSY breaking by boundary condition is a non-local SUSY breaking,
so the resulting soft parameters are more suppressed for
more localized matter fields.
Note that the suppressions of $y_{IJ}$ and $A_{IJ}$
are asymmetric  under $N_I\rightarrow -N_I$.
This is simply because the Yukawa couplings originate from
the boundary at $y=0$.
On the other hand, $m^2_{I\bar{J}}$ are independent of the origin
of the Yukawa couplings, so are symmetric under $N_I\rightarrow -N_I$.

\section{Low energy flavor violations}

In this section, we analyze the low energy flavor
violations resulting from the SS SUSY breaking
for quasi-localized matter fields.
The renormalization scale
for the SS-induced soft parameters of (\ref{radiondomination})
can be identified as the compactification
scale $M_c$. To be specific, here we assume that
$M_c$ is close to the unification scale
$M_{GUT}\sim 2\times 10^{16}$ GeV and
the 4D effective theory below $M_c$ is given by
the MSSM.
For simplicity, we further assume that the two MSSM Higgs doublets $H_1$ and $H_2$
are boundary superfields confined at $y=0$.

Let $\psi_I=\{q_i, u_i, d_i, \ell_i, e_i\}$
($i=1,2,3$) denote the
known three generations of
the left-handed quark-doublets ($q_i$), up-type
antiquark-singlets ($u_i$), down-type antiquark singlets
($d_i$), lepton-doublets ($\ell_i$), and anti-lepton singlets
($e_i$).  The Yukawa
couplings can be written as
\bea
\label{eq:L_Yukawa}
{\cal L}_{\rm Yukawa}=
 y^u_{ij}H_2 u_i q_j+y^d_{ij}H_1 d_i q_j+y^\ell_{ij}H_1 e_i \ell_j
\eea
and the squark/sleptons $\phi_I=\{\tilde{q}_i,\tilde{u}_i, \tilde{d}_i,
\tilde{\ell}_i, \tilde{e}_i\}$
have the soft SUSY breaking couplings:
\bea
\label{eq:L_soft}
{\cal L}_{\rm soft}&=&-\left(\,
 A^u_{ij}H_2\tilde{u}_i\tilde{q}_j+A^d_{ij}H_1\tilde{d}_i\tilde{q}_j
+A^\ell_{ij}H_1\tilde{e}_i\tilde{\ell}_j
\right.\nonumber \\
&&\quad\left.+\,m^{2(\tilde{q})}_{i\bar{j}}\tilde{q}_i
\tilde{q}^*_j
+m^{2(\tilde{u})}_{i\bar{j}}\tilde{u}_i
\tilde{u}_j^*
+m^{2(\tilde{d})}_{i\bar{j}}\tilde{d}_i
\tilde{d}_j^*\right.
\nonumber \\
&&\quad\left.+\,m^{2(\tilde{\ell})}_{i\bar{j}}\tilde{\ell}_i
\tilde{\ell}^*_j
+m^{2(\tilde{e})}_{i\bar{j}}\tilde{e}_i
\tilde{e}_j^*\,\right).
\eea

As shown in (\ref{lambda}), the canonical 4D
Yukawa couplings in (\ref{eq:L_Yukawa}) are
given by
\beq
y^u_{ij} \simeq \lambda^u_{ij}\epsilon^{X^u_i+X^q_j},~~~
y^d_{ij} \simeq  \lambda^d_{ij}\epsilon^{X^d_i+X^q_j},~~~
y^{\ell}_{ij} \simeq  \lambda^{\ell}_{ij}\epsilon^{X^e_i+X^\ell_j},
\label{eq:yukawa}
\eeq
where the redefined  boundary couplings $\lambda^\psi_{ij}$ ($\psi=u,d,\ell$) are
given by
\beq
\lambda^\psi_{ij}=\frac{\tilde{\lambda}^\psi_{ij}\ln(1/\epsilon)}{\pi R}
\sqrt{Z^\psi_iZ^\psi_j}.
\nonumber
\eeq
Here $\tilde{\lambda}^\psi_{ij}$ are the boundary Yukawa couplings which
are generically of order $\sqrt{6\pi^3}/\Lambda_5$ \cite{luty},
\beq
Z_i^\psi=
\left\{
\begin{array}{ll}
|N^\psi_i| & \quad (\,N^\psi_i \neq 0\,) \\
1/[2\ln(1/\epsilon)] & \quad (\,N^\psi_i =0\,)\,,
\end{array}
\right.
\nonumber
\eeq
and $X^\psi_i$ ($\psi=q,u,d,\ell$) are the effective flavor charges defined as
\beq
X^\psi_i =
\left\{
\begin{array}{ll}
N^\psi_i & \quad (\,N^\psi_i\geq 0\,) \\
0 & \quad (\,N^\psi_i< 0\,).
\end{array}
\right.
\nonumber
\eeq
In the following, we assume that $N^\psi_i$ are all integers
and $X^\psi_i$ take the normal hierarchy as
$$X^{\psi}_1 \geq X^{\psi}_2 \geq X^{\psi}_3.
$$
Neglecting the part of $A_{IJ}/y_{IJ}$ suppressed by  $\epsilon^{2|N_{I}|}$
or $\epsilon^{2|N_J|}$ in (\ref{soft1}),
the soft parameters  of (\ref{eq:L_soft})
can be approximated as
\begin{eqnarray}
A^u_{ij} &\simeq & M_0 (X^u_i+X^q_j)\,y^u_{ij},~~
\nonumber \\
A^d_{ij} &\simeq & M_0 (X^d_i+X^q_j)\,y^d_{ij},~~
\nonumber \\
A^\ell_{ij} &\simeq& M_0(X^e_i+X^\ell_j)\,y^\ell_{ij},
\nonumber \\
m^{2(\tilde{\psi})}_{i\bar{j}} &\simeq&
\delta_{i\bar{j}}
M_0^2 \left\{
\begin{array}{ll}
|N^\psi_i|^2 \epsilon^{2|N^\psi_i|}&~~~(\,N^{\psi}_i \neq 0\,)\\
1/[2\ln(1/\epsilon)]^2&~~~(\,N^{\psi}_i = 0\,),
\end{array}
\right.
\label{sssoftmssm}
\end{eqnarray}
where
$$M_0= 2\ln (1/\epsilon)\left(\,\frac{\omega}{R}\,\right)=
-2\mhalf\ln (1/\epsilon)$$
for the universal gaugino mass $\mhalf=-\omega/R$.
The above flavor structure of trilinear $A$-couplings is
 shared by various SUSY--breaking
 scenarios \cite{Kobayashi:2000br},
 while that of soft scalar masses is rather specific to the SS SUSY breaking.

To perform the analysis of low energy observables,
let us first introduce a parameterization of
$\lambda^{\psi}_{ij}$ ($\psi=u,d,\ell$) which are assumed to be of order unity.
As an example, the down-type Yukawa matrix at the unification scale
can be decomposed as
\begin{eqnarray}
y^d_{ij} &=& V_d^\dag \left(
\begin{array}{ccc}
\hat{y}^d_1 & & \\
& \hat{y}^d_2 & \\
&     & \hat{y}^d_3
\end{array}
\right) V_q\,,
\end{eqnarray}
where $V_d, V_q$ are $3\times 3$ unitary matrices
and $\hat{y}^d_i$
can be determined from the observed quark masses.
With this expression, $\lambda^d_{ij}$ can be written as
\begin{eqnarray}
\lambda^d_{ij}
&=&~(\hat{y}^d_1\epsilon^{-X^d_1-X^q_1})
    \,(V_d^*)_{1i}\frac{\epsilon^{X^d_1}}{\epsilon^{X^d_i}}
    \,(V_q)_{1j}\frac{\epsilon^{X^q_1}}{\epsilon^{X^q_j}}
+(\hat{y}^d_2\epsilon^{-X^d_2-X^q_2})
    \,(V_d^*)_{2i}\frac{\epsilon^{X^d_2}}{\epsilon^{X^d_i}}
    \,(V_q)_{2j}\frac{\epsilon^{X^q_2}}{\epsilon^{X^q_j}} \nonumber\\
&&+(\hat{y}^d_3\epsilon^{-X^d_3-X^q_3})
    \,(V_d^*)_{3i}\frac{\epsilon^{X^d_3}}{\epsilon^{X^d_i}}
    \,(V_q)_{3j}\frac{\epsilon^{X^q_3}}{\epsilon^{X^q_j}}.
\label{eq:lambda_d}
\end{eqnarray}
Barring an accidental cancellation between different terms, each term in
 (\ref{eq:lambda_d}) should not exceed ${\cal O}(1)$.
Noting that
$\hat{y}^d_i={\cal O}(\epsilon^{X^d_i+X^q_i})$
and also using the unitarity of the mixing matrices,
we find the following order of magnitude constraints from
 the second and third terms of (\ref{eq:lambda_d}):
\begin{eqnarray}
&&|(V_{d,q})_{21}|\sqrt{1-|(V_{q,d})_{23}|^2}
\lsim \frac{\epsilon^{X^{d,q}_1}}{\epsilon^{X^{d,q}_2}},~~~
\nonumber \\
&& |(V_{d,q})_{31}|\sqrt{1-|(V_{q,d})_{23}|^2}
\lsim \frac{\epsilon^{X^{d,q}_1}}{\epsilon^{X^{d,q}_3}},~~~
\nonumber \\
&& |(V_{d,q})_{32}|\sqrt{1-|(V_{q,d})_{13}|^2}
\lsim \frac{\epsilon^{X^{d,q}_2}}{\epsilon^{X^{d,q}_3}}.
\label{eq:bound}
\end{eqnarray}
Except for the special crossing points with $|(V_{q,d})_{13}| \simeq 1$ or
$|(V_{q,d})_{23}| \simeq 1$,  these constraints are reduced to
\begin{equation}
|(V_{d,q})_{21}| \lsim \frac{\epsilon^{X^{d,q}_1}}{\epsilon^{X^{d,q}_2}},~~~
|(V_{d,q})_{31}| \lsim \frac{\epsilon^{X^{d,q}_1}}{\epsilon^{X^{d,q}_3}},~~~
|(V_{d,q})_{32}| \lsim \frac{\epsilon^{X^{d,q}_2}}{\epsilon^{X^{d,q}_3}}.
\label{eq:bound}
\end{equation}

In general, the unitary matrix $V_d $ can be decomposed
as
\begin{eqnarray}
V_d &=& e^{i\varphi_d} e^{i\vec{\phi}_d^T} \overline{V}_d e^{i\vec{\psi}_d}
,~~~(\sum_i\phi^i_{d}=\sum_i\psi^i_{d}=0)\,,
\label{vd}
\end{eqnarray}
where
\begin{eqnarray}
(\overline{V}_d)_{ij} &=&
\left(
\begin{array}{ccc}
c_{12} & -s_{12} & 0 \\
s_{12} &  c_{12} & 0 \\
0      &  0      & 1
\end{array}
\right)
\left(
\begin{array}{ccc}
c_{13} &  0 & -s_{13}e^{-i\delta^D_{13}} \\
0      &  1 & 0 \\
s_{13}e^{i\delta^D_{13}} &  0      & c_{13}
\end{array}
\right)
\left(
\begin{array}{ccc}
1  & 0 & 0 \\
0 &  c_{23} & -s_{23} \\
0      &  s_{23}      & c_{23}
\end{array}
\right)
\nonumber\\
&=&\left(
\begin{array}{ccc}
c_{12} c_{13} & -s_{12}c_{23}-c_{12}s_{23}s_{13}e^{-i\delta^D_{13}}
& s_{12}s_{23}-c_{12}c_{23}s_{13} e^{-i\delta^D_{13}} \\
s_{12}c_{13}  & c_{12}c_{23}-s_{12}s_{23}s_{13}e^{-i\delta^D_{13}}
& -c_{12}s_{23}-s_{12}c_{23}s_{13}e^{-i\delta^D_{13}} \\
s_{13} e^{i\delta^D_{13}} & s_{23}c_{13} & c_{23}c_{13}
\end{array}
\right)\,,
\label{eq:standard}
\end{eqnarray}
for $s_{ij}\equiv\sin\theta^d_{ij}$ and $c_{ij}\equiv
\cos\theta^d_{ij}$.
With this parameterization, the order of magnitude
constraints (\ref{eq:bound})
 are translated into
\begin{equation}
\theta^d_{12} \lsim \frac{\epsilon^{X^{d}_1}}{\epsilon^{X^{d}_2}},~~~
\theta^d_{13} \lsim \frac{\epsilon^{X^{d}_1}}{\epsilon^{X^{d}_3}},~~~
\theta^d_{23} \lsim \frac{\epsilon^{X^{d}_2}}{\epsilon^{X^{d}_3}}.
\end{equation}
Together with the expression (\ref{vd})
and also a similar expression  of $V_q$, the above
constraints imply
\begin{equation}
|(V_{d,q})_{12}| \lsim \frac{\epsilon^{X^{d,q}_1}}{\epsilon^{X^{d,q}_2}},~~~
|(V_{d,q})_{13}| \lsim \frac{\epsilon^{X^{d,q}_1}}{\epsilon^{X^{d,q}_3}},~~~
|(V_{d,q})_{23}| \lsim \frac{\epsilon^{X^{d,q}_2}}{\epsilon^{X^{d,q}_3}}.
\label{eq:bound2}
\end{equation}

For later discussion, it is convenient to introduce the following
${\cal O}(1)$ parameters:
\begin{equation}
(\kappa_{d,q})_{ij}=\frac{(\overline{V}_{d,q})_{ij}}{
\epsilon^{|X^{d,q}_i-X^{d,q}_j|}}.
\end{equation}
Then using the freedom of rephasing the fields,
we finally arrive at the following parameterizations and
also the constraints, reflecting
well the underlying structure (\ref{eq:yukawa}):
\begin{eqnarray}
&&y^u_{ij} \, =\, \overline{V}^\dag_{u}
\left(
\begin{array}{ccc}
\hat{y}^u_1 e^{i\phi^u_1}& & \\
& \hat{y}^u_2 e^{i\phi^u_2}& \\
&     & \hat{y}^u_3 e^{i\phi^u_3}
\end{array}
\right)
V_{CKM} \overline{V}_q,~~~
\nonumber \\
&& y^d_{ij} \,=\, \overline{V}^\dag_{d}
\left(
\begin{array}{ccc}
\hat{y}^d_1 e^{i\phi^d_1}& & \\
& \hat{y}^d_2 e^{i\phi^d_2}& \\
&     & \hat{y}^d_3 e^{i\phi^d_3}
\end{array}
\right)
\overline{V}_q,
\nonumber\\
&&y^\ell_{ij} \,=\, \overline{V}^\dag_{e}
\left(
\begin{array}{ccc}
\hat{y}^\ell_1 e^{i\phi^\ell_1}& & \\
& \hat{y}^\ell_2 e^{i\phi^\ell_2}& \\
&     & \hat{y}^\ell_3 e^{i\phi^\ell_3}
\end{array}
\right)
\overline{V}_\ell,
\\
&&\theta^\psi_{12} \lsim \frac{\epsilon^{X^{\psi}_1}}{\epsilon^{X^{\psi}_2}},~~~
\theta^\psi_{13} \lsim \frac{\epsilon^{X^{\psi}_1}}{\epsilon^{X^{\psi}_3}},~~~
\theta^\psi_{23} \lsim \frac{\epsilon^{X^{\psi}_2}}{\epsilon^{X^{\psi}_3}},~~~
\nonumber \\
&& \quad\quad (\kappa_{\psi})_{ij} \,\equiv \,
\frac{(\overline{V}_{\psi})_{ij}}{\epsilon^{|X^{\psi}_i-X^{\psi}_j|}}
 \,\,\simeq\,\, {\cal O}(1),
\label{paramet}
\end{eqnarray}
where
$\phi^\psi_1+\phi^\psi_2+\phi^\psi_3=0$
for $\psi=q,u,d,\ell,e$.

Assuming that the CKM matrix does not involve
any fine-tuned cancellation among
mixing angles, we obtain
\begin{equation}
\theta^q_{12} \simeq \epsilon,~~~\theta^q_{23}
\simeq \epsilon^2,~~~\theta^q_{13} \simeq \epsilon^3.
\label{eq:ckm}
\end{equation}
Also the observed fermion mass spectrum indicates
\bea
\frac{\hat{y}^u_1}{\hat{y}^u_3} \simeq 7.2\times 10^{-6} \simeq \epsilon^{7-8}
,&&
\frac{\hat{y}^u_2}{\hat{y}^u_3} \simeq 2.4\times 10^{-3} \simeq \epsilon^{3-4},
\nonumber\\
\frac{\hat{y}^d_1}{\hat{y}^d_3} \simeq 1.0\times 10^{-3} \simeq \epsilon^{4-5}
,&&
\frac{\hat{y}^d_2}{\hat{y}^d_3} \simeq 2.0\times 10^{-2} \simeq \epsilon^{2-3},
\\
\frac{\hat{y}^\ell_1}{\hat{y}^\ell_3} \simeq 2.8\times 10^{-4}
\simeq \epsilon^{5-6},&&
\frac{\hat{y}^\ell_2}{\hat{y}^\ell_3} \simeq 5.9\times 10^{-2} \simeq
\epsilon^{1-2}, \nonumber
\label{eq:yukawa_hierarchy}
\eea
where we assume that $\tan\beta=\langle H_2^0\rangle/\langle H_1^0\rangle$
is not so large. In fact, as for the flavor conserving part,
the SS SUSY breaking with quasi-localized
matter fields is somewhat similar to the gaugino-mediation \cite{gauginomediation} or the
no-scale model since the soft parameters of matter fields
at the compactification scale are suppressed by the quasi-localization
factor. As a result, the model gives either a stau LSP or a negative stau mass-square
when $\tan\beta$ is large \cite{noscale}.
Together with this consideration, $X^\psi_i$ favored by
the observed fermion masses and CKM mixing angles
are given by \cite{cck}
\bea
&&\vec{X}^q = ( 3, 2, 0),~~
\vec{X}^u = ( 4\,\,{\rm or}\,\,5,~ 1\,\,{\rm or}\,\,2,~ 0),~~
\nonumber \\
&& \vec{X}^d = ( 1+x\,\,{\rm or}\,\,2+x,~ x\,\,{\rm or}\,\,1+x,~ x),~~
\nonumber \\
&&\vec{X}^\ell +\vec{X}^e = ( 6+x\,\,{\rm or }\,\,5+x,~ 2+x\,\,
{\rm or}\,\,1+x,~ x),
\label{favored}
\eea
where $x= 1,2$ corresponds to medium and small $\tan\beta$, respectively.

We are now ready to discuss the phenomenology of flavor mixings
resulting from the SS-mediated soft parameters (\ref{sssoftmssm}).
To this end, it is convenient to rotate the fields to the super-CKM basis
in which the Yukawa couplings have diagonal form up to $V_{CKM}$.
In the SCKM basis, the Higgs-slepton trilinear couplings
are given by
\beq
(V_e A^\ell V_\ell^\dag)_{ij} \simeq M_0\left[
 e^{i(\phi^\ell_i-\phi^\ell_j)}\,(\overline{V}_e)_{ik} X^e_k (\overline{V}_e^\dag)_{kj}\,\hat{y}^\ell_{j}
+\hat{y}^\ell_{i}\,(\overline{V}_\ell)_{ik} X^\ell_k (\overline{V}_\ell^\dag)_{kj}\right].
\eeq
Using the unitarity of $V_{e,\ell}$, one easily finds
\begin{eqnarray}
(\overline{V}_e)_{ik} X^e_k (\overline{V}_e^\dag)_{kj} &=&
 (\overline{V}_e)_{i1} (\overline{V}_e^*)_{j1} (X^e_1-X^e_2)
-(\overline{V}_e)_{i3} (\overline{V}_e^*)_{j3} (X^e_2-X^e_3)
+\delta_{ij} X^e_2  \nonumber\\
%(\overline{V}_E)_{ik} X^E_k (\overline{V}_E^\dag)_{kj}
&=& -(\overline{V}_e)_{i2} (\overline{V}_e^*)_{j2} (X^e_1-X^e_2)
-(\overline{V}_e)_{i3} (\overline{V}_e^*)_{j3} (X^e_1-X^e_3)
+\delta_{ij} X^e_1
\end{eqnarray}
and also a similar relation for $X^\ell$ and $\overline{V}_\ell$.
This shows that flavor violation is more suppressed in case that
the lepton superfields in different generation
have a common effective flavor charge.

To parameterize the flavor mixing, let us introduce the
$\tilde{\delta}$ parameters
{\it at the compactification scale}. The $\tilde{\delta}$
parameters for RL/LR mixing are defined as
\begin{equation}
(\tilde{\delta}^{\ell}_{RL})_{ij} = (\tilde{\delta}^{\ell}_{LR})^*_{ji}
 \equiv \frac{(V_e A^\ell V_\ell^\dag)_{ij}}{|\mhalf|^2}v_d,\\
\end{equation}
where $v_d \simeq 174 {\rm GeV} \cos\beta=\langle H_1^0\rangle$.
Using the unitarity relations, we then find
\begin{eqnarray}
(\tilde{\delta}^{\ell}_{RL})_{12}
&\simeq&
-6.44\times10^{-4}\left(\frac{500 {\rm GeV}}{|\mhalf|}\right)
\left(\frac{m_{\mu}(M_{GUT})}{100 {\rm MeV}}\right) \label{eq:deltalG_RL12}
\nonumber\\
&&
\times
\left[
 e^{i(\phi^\ell_1-\phi^\ell_2)}\epsilon^{\Delta X^e_{12}}
\left\{
-\Delta X^e_{12}
+\epsilon^{2\Delta X^e_{23}}~\Delta X^e_{23}
~\frac{(\kappa_e)_{13}(\kappa_e^*)_{23}}{(\kappa_e)_{11}(\kappa_e^*)_{21}}
\right\}
(\kappa_e)_{11}(\kappa_e^*)_{21}
\right.
\nonumber\\
&&
\left.
~~+\left(\frac{m_e}{m_{\mu}}\right)
\epsilon^{\Delta X^\ell_{12}}\left\{
-\Delta X^\ell_{12}
+\epsilon^{2\Delta X^\ell_{23}}~\Delta X^\ell_{23}
~\frac{(\kappa_\ell)_{13}(\kappa_\ell^*)_{23}}{(\kappa_\ell)_{11}(\kappa_\ell^*)_{21}}
\right\}
(\kappa_\ell)_{11}(\kappa_\ell^*)_{21}
\right],
\nonumber
\\
%\end{eqnarray}
%\begin{eqnarray}
(\tilde{\delta}^{\ell}_{RL})_{13} &\simeq&
-6.44\times 10^{-3} \left(\frac{500 {\rm GeV}}{|\mhalf|}\right)
\left(\frac{m_\tau(M_{GUT})}{1.00 {\rm GeV}}\right)
\nonumber\\
&&\times \left[
e^{i(\phi^\ell_1-\phi^\ell_3)} \epsilon^{\Delta X^e_{13}}
\left\{
\Delta X^e_{13}+\frac{(\kappa_e)_{12}(\kappa_e^*)_{32}}{(\kappa_e)_{13}
(\kappa_e^*)_{33}}\Delta X^e_{12}
\right\}(\kappa_e)_{13}(\kappa_e^*)_{33}
\right.
\nonumber\\
&&
\phantom{\times}
\left.
+\left(\frac{m_e}{m_\tau}\right)
\epsilon^{\Delta X^\ell_{13}}
\left\{
\Delta X^\ell_{13}+\frac{(\kappa_\ell)_{12}(\kappa_\ell^*)_{32}}{(\kappa_\ell)_{13}(\kappa_\ell^*)_{33}}\Delta X^\ell_{12}
\right\}(\kappa_\ell)_{13}(\kappa_\ell^*)_{33}
\right],
\nonumber\\
%\end{eqnarray}
%\begin{eqnarray}
(\tilde{\delta}^{\ell}_{RL})_{23} &\simeq&
-6.44\times 10^{-3} \left(\frac{500 {\rm GeV}}{|\mhalf|}\right)
\left(\frac{m_\tau(M_{GUT})}{1.00 {\rm GeV}}\right)\nonumber\\
&&
\times
\left[
e^{i(\phi^\ell_2-\phi^\ell_3)}
\epsilon^{\Delta X^e_{23}}
\left\{
  \Delta X^e_{23}
  -\epsilon^{2\Delta X^e_{12}}
   \frac{(\kappa_e)_{21}(\kappa_e^*)_{31}}{(\kappa_e)_{23}(\kappa_e^*)_{33}}
  \Delta X^e_{12}
\right\}(\kappa_e)_{23}(\kappa_e^*)_{33}
\right.
\nonumber\\
&&
\left.
\phantom{\times}
+\left(\frac{m_{\mu}}{m_{\tau}}\right)
\epsilon^{\Delta X^\ell_{23}}
\left\{
  \Delta X^\ell_{23}
  -\epsilon^{2\Delta X^\ell_{12}}
   \frac{(\kappa_\ell)_{21}(\kappa_\ell^*)_{31}}{(\kappa_\ell)_{23}(\kappa_\ell^*)_{33}}  \Delta X^\ell_{12}
\right\}
(\kappa_\ell)_{23}(\kappa_\ell^*)_{33}
\right],
\end{eqnarray}
where $(\kappa_\psi)_{ij}={\cal O}(1)$
as defined in (\ref{paramet}) and
$\Delta X^{\psi}_{ij} \equiv X^{\psi}_i - X^{\psi}_j.$

Note that if $X^e_1 = X^e_2$ or $X^e_2=X^e_3$,
$(\tilde{\delta}^{\ell}_{RL})_{12}$ or
$(\tilde{\delta}^{\ell}_{RL})_{23}$ receives a
suppression by $\epsilon^{2\Delta X^e_{23}}$ or
$\epsilon^{2\Delta X^e_{12}}$, respectively.
If $X^e_1=X^e_2=X^e_3$, all contributions to
$(\tilde{\delta}^\ell_{RL})_{12,13,23}$
from the right-handed sleptons disappear,  leaving only
the left-handed slepton contributions which are
suppressed by $(m_e/m_{\mu})$, $(m_e/m_{\tau})$ and $(m_\mu/m_{\tau})$, respectively.
Since $X^\psi_i$ are quantized, this suppression mechanism
does not require any fine tuning of parameters.
We stress that this suppression of the flavor violations from
$A_{IJ}/y_{IJ}$  relies on the specific feature of the SS SUSY breaking
that $A_{IJ}/y_{IJ}$ at the compactification scale are quantized as in
 (\ref{soft1}) if the kink masses are quantized.
Expressions for $(\tilde{\delta}^{\ell}_{RL})_{21,31,32}$ can be obtained
from $(\tilde{\delta}^{\ell}_{RL})_{12,13,23}$ by
exchanging $e\leftrightarrow\ell$ for
$\Delta X^{e,\ell}$ and $\kappa_{e,\ell}$
together with the exchange $1\leftrightarrow 2$, $1\leftrightarrow 3$ and $2\leftrightarrow 3$,
respectively for
the first index $i$ in $(\kappa_{e,\ell})_{ij}$,
and also moving the phase factor in the first line of the rectangular
parenthesis to the second line. A parallel discussion for
 $(\tilde{\delta}^d_{RL})_{ij}$ and $(\tilde{\delta}^u_{RL})_{ij}$
can be easily understood, so we do not repeat it here.

In the SCKM basis, the right-handed down-type squark masses are given by
\bea
(V_d m^{2(\tilde{d})} V_d^\dag)_{i\bar{j}} &\simeq& M_0^2\,
 e^{i(\phi^d_i-\phi^d_j)}\,
(\overline{V}_d)_{ik}|N^d_k|^2 \epsilon^{2|N^d_k|}(\overline{V}_d^\dag)_{kj}.
%(V_q m^{2(q)} V_q^\dag)_{i\bar{j}} &\simeq& (2\ln 5)^2
%(\overline{V}_q)_{ik}(X^q_k)^2 \epsilon^{2X^q_k}(\overline{V}_q^\dag)_{kj}
\eea
The corresponding $\tilde{\delta}$ parameters
at the compactification scale can be defined as
\beq
(\tilde{\delta}^{d}_{RR})_{ij}
 \equiv \frac{(V_d m^{2(\tilde{d})} V_d^\dag)_{ij}}{|\mhalf|^2}.
\eeq
We then find
\bea
\label{eq:deltad_RR12}
(\tilde{\delta}^{d}_{RR})_{12}
&\simeq&
-10 e^{i(\phi^d_1-\phi^d_2)} \epsilon^{\Delta X^d_{12}}
(\kappa_d)_{11}(\kappa_d)^*_{21}
\nonumber\\
&&
\phantom{-10}
\times
\left[\left(|N^d_2|^2\epsilon^{2 |N^d_2|}
        -|N^d_1|^2\epsilon^{2 |N^d_1|}\right)
-\frac{(\kappa_d)_{13}(\kappa_d)^*_{23}}
               {(\kappa_d)_{11}(\kappa_d)^*_{21}}\epsilon^{2\Delta X^d_{23}}
              \left( |N^d_3|^2 \epsilon^{2|N^d_3|}-|N^d_2|^2 \epsilon^{2|N^d_2|} \right)\right],
\nonumber \\
\label{eq:deltad_RR13}
(\tilde{\delta}^{d}_{RR})_{13} &\simeq&
10 e^{i(\phi^d_1-\phi^d_3)} \epsilon^{\Delta X^d_{13}}
(\kappa_d)_{13}(\kappa_d)^*_{33}
\nonumber\\
&&
\phantom{10}
\times
\left[\left(|N^d_3|^2\epsilon^{2|N^d_3|}-|N^d_1|^2\epsilon^{2|N^d_1|}
\right)
+\frac{(\kappa_d)_{12}(\kappa_d)^*_{32}}
               {(\kappa_d)_{13}(\kappa_d)^*_{33}}
              \left( |N^d_2|^2\epsilon^{2|N^d_2|}
                -|N^d_1|^2\epsilon^{2|N^d_1|}\right)
              \right],
\nonumber \\
\label{eq:deltad_RR23}
(\tilde{\delta}^{d}_{RR})_{23} &\simeq&
 10 e^{i(\phi^d_2-\phi^d_3)} \epsilon^{\Delta X^d_{23}}
(\kappa_d)_{12}(\kappa_d)^*_{22}
\nonumber\\
&&
\phantom{10}
\times
\left[\left(|N^d_3|^2\epsilon^{2|N^d_3|}-|N^d_2|^2\epsilon^{2|N^d_2|}\right)
-\frac{(\kappa_d)_{21}(\kappa_d)^*_{31}}
               {(\kappa_d)_{23}(\kappa_d)^*_{33}}\epsilon^{2\Delta X^d_{12}}
              \left(
               |N^d_2|^2\epsilon^{2|N^d_2|}-|N^d_1|^2\epsilon^{2|N^d_1|}\right)\right],
\eea
where the factor $|N^d_i|^2\epsilon^{2|N^d_i|}$ should be replaced by
$1/[2\ln(1/\epsilon)]^2$ for $N^d_i=0$.
A suppression of the flavor violations from $m^{2(\tilde{d})}_{i\bar{j}}$ is possible if
any pair of $|N^d_i|$ have a common value.
In particular, all flavor mixings disappear if
$|N^d_1|=|N^d_2|=|N^d_3|$, although this is not favored by the observed
quark masses and CKM mixing angles as was noted in the discussion
leading to (\ref{favored}).
Again this suppression of the flavor violations from $m^2_{I\bar{J}}$ relies
on the specific feature of the SS SUSY breaking that $m^2_{I\bar{J}}$ at the
compactification scale are quantized as in (\ref{soft1}) if the kink masses are
quantized.
Expressions of $(\tilde{\delta}^{\ell}_{RR})$ and $(\tilde{\delta}^{u}_{RR})$
can be obtained by appropriately changing
the flavor indices.
To obtain  $(\tilde{\delta}^{\ell}_{LL})$, $(\tilde{\delta}^{d}_{LL})$ and
$(\tilde{\delta}^{u}_{LL})$, one has to remove the
phase factors in addition to the necessary change of flavor indices.

The soft terms and mixing parameters discussed above are given at the
compactification scale, thus a
renormalization group (RG) improvement is required.
One approach is a full numerical calculation including
the effects of flavor mixing in the RG evolution.
This approach would be necessary
 when the leading flavor mixing comes from the loop effects \cite{radiative_gut,Barbieri:1995tw,seesaw,Hisano:1998fj,seesaw_gut}.
However in our case, it is not so useful since the model involves many
free parameters.
We thus use an approximate analytic solution
ignoring  the RG effects on the flavor off-diagonal part of the soft
parameters \cite{analytic}. This approximation is reasonably good for
the order of magnitude  analysis of low energy flavor violations.
In the same spirit, we use the mass-insertion approximation
\cite{massinsertion}
to calculate flavor-violating observables at the weak scale,
rather than using the more accurate mass-eigenstate formalism.

Using for instance the results of \cite{Barbieri:1995tw}, we find that
the gaugino masses $M_a$
and the flavor-diagonal sfermion masses
$m^{2(\tilde{\psi})}_{i\bar{i}}$  at the weak scale are given by
\begin{eqnarray}
|M_a^2|/|\mhalf|^2 &=& 0.16\,:\, 0.67\,:\, 8.5
\nonumber \\
m^{2(\tilde{q})}_{i\bar{i}}/|\mhalf|^2&\simeq & 7.2\,:\, 7.2\,:\, 6.0
\nonumber \\
m^{2(\tilde{u})}_{i\bar{i}}/|\mhalf|^2&\simeq & 6.7\,:\, 6.7\,:\, 4.8
\nonumber \\
m^{2(\tilde{d})}_{i\bar{i}}/|\mhalf|^2&\simeq & 6.7 \quad (i=1,2,3)
\nonumber \\
m^{2(\tilde{\ell})}_{i\bar{i}}/|\mhalf|^2&\simeq & 0.53 \quad (i=1,2,3)
\nonumber \\
m^{2(\tilde{e})}_{i\bar{i}}/|\mhalf|^2&\simeq & 0.15 \quad (i=1,2,3),
\label{rg}
\end{eqnarray}
where $\mhalf=-\omega/R$  is the universal gaugino mass at
the compactification scale and $a=1,2,3$ stand
for $U(1)_Y\times SU(2)_L\times SU(3)_c$, respectively.
To get these numerical results, we used the top quark Yukawa
coupling $y_t\sim 1$ at the weak scale and ignored
all other Yukawa couplings.
Similarly the Higgs masses at the weak scale are found to be
\begin{equation}
m^2_{H_1}/|\mhalf|^2
\simeq 0.53 \,,\quad
m^2_{H_2}/|\mhalf|^2\simeq -3.0.
\end{equation}
We also approximate the Higgsino mass parameter $\mu$ as
\begin{eqnarray}
\mu^2 &=& -\frac{M_Z^2}{2}-\frac{m^2_{H_1}-m^2_{H_2}\tan^2\beta}{1-\tan^2\beta}\nonumber\\
      &\simeq& -\frac{M_Z^2}{2} +
\left(3.0-\frac{3.5}{1-\tan^2\beta}\right)|\mhalf|^2 \,
\simeq \, 3|\mhalf|^2.
\label{eq:mu-term}
\end{eqnarray}
%
%In fact,
%above no-scale type spectrum is only valid for small $\tan\beta$ and
%large $\tan\beta$ solution does not exist because stau develops unstable
% vacuum.

The slepton $\delta$ parameters at the weak scale are defined
 in the SCKM basis as \cite{massinsertion},
\beq
(\delta^\ell_{RL})_{ij} \equiv \frac{A^\ell_{ij} v_d-\mu^*m^\ell_i\delta_{ij}\tan\beta}{\sqrt{m^{2(\tilde{e})}_{i\bar{i}} m^{2(\tilde{\ell})}_{j\bar{j}}}},~~
%\\
%&&
(\delta^\ell_{RR})_{ij} \equiv  \frac{m^{2(\tilde{e})}_{i\bar{j}}}
{\sqrt{m^{2(\tilde{e})}_{i\bar{i}} m^{2(\tilde{e})}_{j\bar{j}}}},~~
%\\
(\delta^\ell_{LL})_{ij} \equiv
\frac{m^{2(\tilde{\ell})}_{i\bar{j}}}{\sqrt{m^{2(\tilde{\ell})}_{i\bar{i}}
m^{2(\tilde{\ell})}_{j\bar{j}}}}.
\eeq
According to (\ref{rg}), these weak scale
$\delta$ parameters are related to the GUT scale $\tilde{\delta}$
parameters as
\beq
(\delta^\ell_{RL})_{ij} \simeq 3.5\, (\tilde{\delta}^{\ell}_{RL})_{ij},~~
(\delta^\ell_{RR})_{ij} \simeq 6.7\, (\tilde{\delta}^{\ell}_{RR})_{ij},~~
(\delta^\ell_{LL})_{ij} \simeq 1.9\,
(\tilde{\delta}^{\ell}_{LL})_{ij}~~~(i \neq j),
\label{eq:deltal_RL12}
\eeq
and similar expressions can be obtained also for the squark
$\delta$ parameters.

It turns  out that the most dangerous low energy flavor
violations in our SS SUSY breaking scenario
are $\mu\rightarrow e\gamma$ and $\epsilon_K$.
Let us first discuss the $\mu \to
e\gamma$ process.
Neglecting the electron mass in the final state, the
$\mu\to e\gamma$ branching ratio
is given by the sum of two branching ratios
with opposite chirality,
\begin{eqnarray}
BR(\mu^+ \to e^+\gamma) &\simeq& BR(\mu^+_L \to e^+_R \gamma)
                            +BR(\mu^+_R \to e^+_L \gamma).
\end{eqnarray}
Assuming the sparticle spectrum of (\ref{rg})
and (\ref{eq:mu-term}), we find
\begin{eqnarray}
\left[\frac{BR(\mu^+_R \to e^+_L \gamma)}{1.2\times10^{-11}}
\right]^{1/2}
&\simeq&
\left( \frac{500{\rm GeV}}{|\mhalf|}\right)^2
\nonumber\\
&&\times
\left|-\frac{(\delta^\ell_{RR})_{12}}{3.1\times10^{-2}
              /(e^{-i\theta_\mu}\tan\beta+2.7)}
%\right.
%\nonumber\\
%&&
%\phantom{\times |}
+\frac{(\delta^\ell_{RR})_{13}(\delta^\ell_{RR})_{32}}{5.1\times10^{-2}
              /(e^{-i\theta_\mu}\tan\beta+3.0)}
\right.
\nonumber\\
&&
\phantom{\times|}
+\left( \frac{|\mhalf|}{500{\rm GeV}}\right)\left( \frac{106{\rm MeV}}{m_\mu}\right)
\nonumber\\
&&
\phantom{++}
\times
\left\{
\frac{(\delta^\ell_{RL})_{12}}{4.8\times10^{-6}}
%%\right.
%\nonumber\\
%&&
%\phantom{\times|}
-
%\left( \frac{|\mhalf|}{500{\rm GeV}}\right)\left( \frac{106{\rm MeV}}{m_\mu}\right)
\frac{(\delta^\ell_{RL})_{11}(\delta^\ell_{LL})_{12}+(\delta^\ell_{RL})_{13}(\delta^\ell_{LL})_{32}}{6.2\times10^{-6}}
\right.
\nonumber\\
&&\phantom{\times|}
\phantom{++\times}
-
\left.\left.
%\left( \frac{|\mhalf|}{500{\rm GeV}}\right)\left( \frac{106{\rm MeV}}{m_\mu}\right)
\frac{(\delta^\ell_{RR})_{12}(\delta^\ell_{RL})_{22}+(\delta^\ell_{RR})_{13}(\delta^\ell_{RL})_{32}}{8.8\times10^{-6}}
%\nonumber\\
%&&\left.
%\phantom{\times|}
+
%\left( \frac{|\mhalf|}{500{\rm GeV}}\right)\left( \frac{106{\rm MeV}}{m_\mu}\right)
\frac{(\delta^\ell_{RR})_{13}(\delta^\ell_{RL})_{33}(\delta^\ell_{LL})_{32}}
{1.1\times10^{-5}}
\right\}
\right|,
\label{eq:muReLg}
\end{eqnarray}
\begin{eqnarray}
\left[\frac{BR(\mu^+_L \to e^+_R \gamma)}{1.2\times10^{-11}}
\right]^{1/2}
&\simeq&
\left( \frac{500{\rm GeV}}{|\mhalf|}\right)^2\nonumber\\
&&
\times\left|\frac{(\delta^\ell_{LL})_{12}}{2.6\times10^{-2}
              /(e^{i\theta_\mu}\tan\beta+0.11)}
%\right.
%\nonumber\\
%&&
%\phantom{\times|}
-\frac{(\delta^\ell_{LL})_{13}(\delta^\ell_{LL})_{32}}{7.8\times10^{-2}
              /(e^{i\theta_\mu}\tan\beta-0.42)}
\right.
\nonumber\\
&&\phantom{\times|}+\left( \frac{|\mhalf|}{500{\rm GeV}}\right)\left( \frac{106{\rm MeV}}{m_\mu}\right)
\nonumber\\
&&
\phantom{++}
\times\left\{\frac{(\delta^\ell_{LR})_{12}}{4.8\times10^{-6}}
%\nonumber\\
%&&
%\phantom{\times|}
-
%\left( \frac{|\mhalf|}{500{\rm GeV}}\right)\left( \frac{106{\rm MeV}}{m_\mu}\right)
\frac{(\delta^\ell_{LR})_{11}(\delta^\ell_{RR})_{12}+(\delta^\ell_{LR})_{13}(\delta^\ell_{RR})_{32}}{8.8\times10^{-6}}
\right.
\nonumber\\
&&\phantom{\times|}
\phantom{++\times}
-
\left.\left.
%\left( \frac{|\mhalf|}{500{\rm GeV}}\right)\left( \frac{106{\rm MeV}}{m_\mu}\right)
\frac{(\delta^\ell_{LL})_{12}(\delta^\ell_{LR})_{22}+(\delta^\ell_{LL})_{13}(\delta^\ell_{LR})_{32}}{6.2\times10^{-6}}
%\nonumber\\
%&&\left.\phantom{\times|}
+
%\left( \frac{|\mhalf|}{500{\rm GeV}}\right)\left( \frac{106{\rm MeV}}{m_\mu}\right)
\frac{(\delta^\ell_{LL})_{13}(\delta^\ell_{LR})_{33}(\delta^\ell_{RR})_{32}}
{1.1\times10^{-5}}
\right\}
\right|,
\label{eq:muLeRg}
\end{eqnarray}
where
$\theta_\mu \equiv {\rm arg}(\mu \mhalf)$
 and the branching ratio is divided by the latest upperbound
 $BR(\mu \to e\gamma)<1.2 \times 10^{-11}$ \cite{Brooks:1999pu}.
Corresponding analytic formulas can be found {\it e.g.} in \cite{Hisano:1998fj}
 and we expanded chargino and neutralino mixings up to ${\cal
O}(M_{W,Z}/|\mhalf|)$. Here we include two insertions of $\delta$ for
the $RR$ and $LL$ channels, while only a single insertion
of $\delta$ is included for the $RL$ channel.
Note that not only $(2,1)$ mixings but also some
combinations of $(2,3)$ and $(3,1)$ mixings
are severely constrained by $\mu\to e\gamma$.
Similar expressions for $BR(\tau \to e \gamma)$
 and $BR(\tau \to \mu \gamma)$ can be obtained by
 replacing $m_\mu$ with $m_\tau$,
 changing the generation indices in the $\delta$ parameters as $2
 \leftrightarrow 3$ and $1 \to 2 \to 3 \to 1$, and
multiplying $BR(\tau\to e\nu_\tau\bar{\nu}_e)=0.178$
and $BR(\tau\to \mu\nu_\tau\bar{\nu}_\mu)=0.174$ %\cite{Hagiwara:fs}
, respectively.

Eqs.(\ref{eq:deltalG_RL12})
and (\ref{eq:deltal_RL12}) suggest that if none of
$\Delta X^{e,\ell}_{12}$ vanishes,
 $|(\delta^\ell_{RL})_{12,21}|={\cal O}(10^{-3}\epsilon^{\Delta
X^{e,\ell}_{12}})$ for $|\mhalf|\sim 500$ GeV, so it is difficult to satisfy
the experimental bound
$|(\delta^\ell_{RL})_{12,21}| \lsim {\cal O}(10^{-6})$
under the constraint $(\Delta X^e)_{12}+(\Delta X^\ell)_{12}=3$,
$4$ or $5$ which comes from $m_e/m_\mu$.
A simple mechanism to suppress the $\mu\to e\gamma$ rate is to
choose some of the quantized lepton kink masses to be degenerate.
For instance, if
$X^\ell_1=X^\ell_2=X^\ell_3$ or $X^e_1=X^e_2=X^e_3$,
the $\mu\to e\gamma$ bound can be safely satisfied without
severe fine-tuning of the involved ${\cal O}(1)$ parameters for
$|\mhalf|=500$ GeV.
Table \ref{table:muegconst} summarizes the possible choices of
the flavor charge differences which reproduce the correct charged lepton mass
spectrum. The resulting
$\mu \to e\gamma$ rate expressed in terms
of $(\delta^\ell_{RL})_{12,21}$ for $|\mhalf|=500 {\rm GeV}$.
For this, we set all ${\cal O}(1)$ parameters, i.e. $\kappa_{e,\ell}$,
to be unity, so the results of Table \ref{table:muegconst} should be
interpreted as a kind of order of magnitude estimate.
A double check in the table indicates that the model can safely satisfy
the $\mu\to e\gamma$ constraint without any fine tuning of parameters,
and a single check means one may need a mild tuning of
parameters.
%This result clearly shows that the $\mu\to e\gamma$
%constraint favors the models in which two of $(\Delta X^e)_{ij}$ and
%$(\Delta X^{\ell})_{ij}$ ($j>i$) vanish.

Tables \ref{table:lfv2}$-$\ref{table:lfv4} represent
the lepton flavor violating rates predicted by
the models of Table
 \ref{table:muegconst}.
We used Eqs.(\ref{eq:muReLg}) and (\ref{eq:muLeRg})
with $|\mhalf|=500\, {\rm GeV}$ for $\mu\to e\gamma$,
and the analogous formula for $\tau\to\mu\gamma$ or $e\gamma$.
In this procedure, we set all the involved ${\cal O}(1)$ parameters to unity, and
added  all contributions constructively.
For $(\delta^\ell_{RL})_{ii}$, we include $\tan\beta$ enhanced
 F-term contribution,
\beq
(\delta^\ell_{RL})_{ii}\simeq -
 0.012e^{-i\theta_\mu} \tan\beta\, \left(m^\ell_i/1 {\rm GeV}\right)\left(500 {\rm GeV}/|\mhalf|\right).
\eeq
Taking into account the sensitivity of next generation experiments
\cite{meg,LOI}, the $\mu\to e\gamma$ ($\tau\to e\gamma$ or $\mu \gamma$)
branching ratio smaller than $10^{-14}$ ($10^{-9}$) for
$\tan\beta = 10$ is not depicted in
the Tables.
Models indicated by light color lead to a too rapid $\mu\to e\gamma$
even when the ${\cal O}(1)$ parameters are assumed to suppressed
by a factor of $1/4$ for $|\mhalf|=500$ GeV.
Note that all branching ratios scale as $|\mhalf|^{-4}$, thus the numbers
in Tables \ref{table:lfv2}$-$\ref{table:lfv4} decrease (increase) by
a factor of $1/16$ ($16$) when $|\mhalf|=1$ TeV ($250$ GeV).
Many of the models in Tables \ref{table:lfv2} and \ref{table:lfv4} lead to
$\ell_i\to\ell_j\gamma$ which
can be explored by future experiments for a reasonable range
of ${\cal O}(1)$ parameters.
Some models already start to overlap with the latest
 bound $BR(\tau \to \mu\gamma)<3.1\times 10^{-7}$ \cite{Abe:2003sx}
 with an ambiguity associated with ${\cal O}(1)$ parameters.
Different choices of the effective flavor charges
predict different patterns of lepton flavor violation,
thus a  combinatoric analysis of different experiments will
be useful for distinguishing models discussed here.
In particular,  determining the chirality pattern of the processes can provide
a crucial information on the model \cite{chirality}.
Note that the chirality pattern for $X^{\ell}_1=X^\ell_2=X^\ell_3$
is opposite to the case with $X^e_1=X^e_2=X^e_3$
which has the same chirality pattern as the
lepton flavor violating decays in seesaw models
\cite{seesaw,Hisano:1998fj,muegpol}.

\begin{table}[h]
\begin{center}
\begin{tabular}{|c|c|c|c|c|c|c|c|c|c|}
\hline
\multicolumn{5}{|c|}{$\hat{y}^\ell_i/\hat{y}^\ell_3={\cal O}(\epsilon^5,\epsilon,1)$}&
\multicolumn{5}{|c|}{$\hat{y}^\ell_i/\hat{y}^\ell_3={\cal O}(\epsilon^6,\epsilon,1)$}\\
\hline
$\Delta X^e_{i3}$ & $\Delta X^\ell_{i3}$ &
 $\frac{|(\delta^\ell_{RL})_{12}|}{4.8\times10^{-6}}$ &
 $\frac{|(\delta^\ell_{LR})_{12}|}{4.8\times10^{-6}}$ &
&
$\Delta X^e_{i3}$ & $\Delta X^\ell_{i3}$ &
 $\frac{|(\delta^\ell_{RL})_{12}|}{4.8\times10^{-6}}$ &
 $\frac{|(\delta^\ell_{LR})_{12}|}{4.8\times10^{-6}}$ &
\\
\hline
$-$      &    $-$   & $-$ & $-$  &
                   & $(6,1,0)$&$(0,0,0)$ & 0.80 & 0.040&$\sqrt{}$$\sqrt{}$\\
$(5,1,0)$&$(0,0,0)$ & 3.2& 0.015&$\sqrt{}$
                    & $(5,1,0)$&$(1,0,0)$ & 3.2  & 100   &\\
$(4,1,0)$&$(1,0,0)$ & 12 & 100   &
                    & $(4,1,0)$&$(2,0,0)$ & 12  & 40  &\\
$(3,1,0)$&$(2,0,0)$ & 40  & 40  &
                    & $(3,1,0)$&$(3,0,0)$ & 40   & 12  &\\
$(2,1,0)$&$(3,0,0)$ & 100  & 12  &
                    & $(2,1,0)$&$(4,0,0)$ & 100   & 3.2 &\\
$(1,1,0)$&$(4,0,0)$ & 20 & 3.2 &
                    & $(1,1,0)$&$(5,0,0)$ & 20  & 0.80 &\\
  $-$    &    $-$   & $-$ & $-$  &
                    & $(5,0,0)$&$(1,1,0)$ & 0.80 & 20  &\\
$(4,0,0)$&$(1,1,0)$ & 3.2 & 20 &
                    & $(4,0,0)$&$(2,1,0)$ & 3.2  & 100   &\\
$(3,0,0)$&$(2,1,0)$ & 12 & 100   &
                    & $(3,0,0)$&$(3,1,0)$ & 12  & 40  &\\
$(2,0,0)$&$(3,1,0)$ & 40  & 40  &
                    & $(2,0,0)$&$(4,1,0)$ & 40   & 12  &\\
$(1,0,0)$&$(4,1,0)$ & 100  & 12  &
                    & $(1,0,0)$&$(5,1,0)$ & 100   & 3.2 &\\
$(0,0,0)$&$(5,1,0)$ & 0.015 & 3.2 &$\sqrt{}$
                    & $(0,0,0)$&$(6,1,0)$ & 0.040& 0.80&$\sqrt{}$$\sqrt{}$\\
\hline
%\end{tabular}
%\caption{\label{table:muegconst1}}
%\end{center}
%\end{table}
%
%\begin{table}[h]
%\begin{center}
%\begin{tabular}{|c|c|c|c|c|c|c|c|c|c|}
\hline
\multicolumn{5}{|c|}{$\hat{y}^\ell_i/\hat{y}^\ell_3={\cal O}(\epsilon^5,\epsilon^2,1)$}&
\multicolumn{5}{|c|}{$\hat{y}^\ell_i/\hat{y}^\ell_3={\cal O}(\epsilon^6,\epsilon^2,1)$}\\
\hline
$\Delta X^e_{i3}$ & $\Delta X^\ell_{i3}$ &
$\frac{|(\delta^l_{RL})_{12}|}{4.8\times10^{-6}}$ &
$\frac{|(\delta^l_{LR})_{12}|}{4.8\times10^{-6}}$ &
&
$\Delta X^e_{i3}$ & $\Delta X^\ell_{i3}$ &
$\frac{|(\delta^l_{RL})_{12}|}{4.8\times10^{-6}}$ &
$\frac{|(\delta^l_{LR})_{12}|}{4.8\times10^{-6}}$ &
\\
\hline
   $-$   &   $-$   & $-$ & $-$   &
                   &$(6,2,0)$&$(0,0,0)$& 3.2 & 0.60 &$\sqrt{}$\\
$(5,2,0)$&$(0,0,0)$& 12 & 0.057 &
                   &$(5,2,0)$&$(1,0,0)$& 12 & 100    &\\
$(4,2,0)$&$(1,0,0)$& 40  & 100    &
                   &$(4,2,0)$&$(2,0,0)$& 40  & 40   &\\
$(3,2,0)$&$(2,0,0)$& 100  & 40   &
                   &$(3,2,0)$&$(3,0,0)$& 100  & 12   &\\
$(2,2,0)$&$(3,0,0)$& 1.6  & 12&
                   &$(2,2,0)$&$(4,0,0)$& 1.6& 3.2&$\sqrt{}$\\
$-$      &$-$      & $-$ & $-$   &
                   &$(5,1,0)$&$(1,1,0)$& 3.2 & 20   &\\
$(4,1,0)$&$(1,1,0)$& 12 & 20  &
                   &$(4,1,0)$&$(2,1,0)$& 12 & 100    &\\
$(3,1,0)$&$(2,1,0)$& 40  & 100    &
                   &$(3,1,0)$&$(3,1,0)$& 40  & 40   &\\
$(2,1,0)$&$(3,1,0)$& 100  & 40   &
                   &$(2,1,0)$&$(4,1,0)$& 100  & 12   &\\
$(1,1,0)$&$(4,1,0)$& 20 & 12   &
                   &$(1,1,0)$&$(5,1,0)$& 20 & 3.2  &\\
$-$      & $-$     & $-$ &  $-$  &
                   &$(4,0,0)$&$(2,2,0)$& 3.2 & 1.6  &$\sqrt{}$\\
$(3,0,0)$&$(2,2,0)$& 12 & 1.6 &
                   &$(3,0,0)$&$(3,2,0)$& 12 & 100    &\\
$(2,0,0)$&$(3,2,0)$& 40  & 100    &
                   &$(2,0,0)$&$(4,2,0)$& 40 & 40   &\\
$(1,0,0)$&$(4,2,0)$& 100  & 40    &
                   &$(1,0,0)$&$(5,2,0)$& 100  & 12   &\\
$(0,0,0)$&$(5,2,0)$& 0.057  & 12
                   & &$(0,0,0)$&$(6,2,0)$& 0.60 &3.2&$\sqrt{}$\\
\hline
\end{tabular}
\caption{ Lepton mass hierarchy vs constraint from $\mu \to e\gamma$.
Here $(\delta^\ell_{RL})_{12,21}$ is divided by the values saturating $BR(\mu
 \to e\gamma)= 1.2\times 10^{-11}$ for $|\mhalf|=500 {\rm GeV}$.
A double check indicates that the model can safely satisfy the $\mu \to
 e\gamma$ constraint and a single check means one may need a mild tuning of
 ${\cal O}(1)$ parameters.
 \label{table:muegconst}}
\end{center}
\end{table}

\begin{table}
\begin{center}
\begin{tabular}{|c|c|c|c|c|}
\hline
\multicolumn{5}{|c|}{$y^\ell_i={\cal O}(\epsilon^7,\epsilon^3,\epsilon^2)$}\\
\hline
$N^e_i$ & $N^\ell_i$ &
$BR(\mu^+_R \to e^+_L,\gamma)$ &
$BR(\tau^+_R \to e^+_L,\gamma)$ &
$BR(\tau^+_R \to \mu^+_L,\gamma)$
\\
\hline
$(5,1,-2)$&$(2,2,2)$
 &$1.5(1+0.14 t_\beta)^2\times 10^{-10}$
 & $-$
 & $2.4(1+0.091t_\beta)^2\times 10^{-8}$\\
$(5,1,-1)$&$(2,2,2)$
 &$1.6(1+0.14 t_\beta)^2\times 10^{-10}$
 & $-$
 & $2.1\times 10^{-8}$\\
$(6,2,1)$&$(1,1,1)$
 &$1.3(1+0.026 t_\beta)^2\times 10^{-10}$
 & $-$
 & $4.4(1+0.42t_\beta)^2\times 10^{-8}$\\
$(7,3,2)$&$(-1,-1,-1)$
 &$1.2\times 10^{-10}$
 & $-$
 & $2.4(1+0.091t_\beta)^2\times 10^{-8}$\\
\hline
$(-1,-1,-1)$&$(7,3,2)$
 &$-$
 &$-$
 &$-$\\
$(1,1,1)$&$(6,2,1)$
 &$-$
 & $-$
 &$-$\\
$(2,2,2)$&$(5,1,-1)$
 &$-$
 &$-$
 &$-$\\
$(2,2,2)$&$(5,1,-2)$
 &$-$
 &$-$
 &$-$\\
\hline
$N^e_i$ & $N^\ell_i$ &
$BR(\mu^+_L \to e^+_R,\gamma)$ &
$BR(\tau^+_L \to e^+_R,\gamma)$ &
$BR(\tau^+_L \to \mu^+_R,\gamma)$
\\
\hline
$(5,1,-2)$&$(2,2,2)$
 &$-$
 &$-$
 &$-$\\
$(5,1,-1)$&$(2,2,2)$
 &$-$
 &$-$
 &$-$\\
$(6,2,1)$&$(1,1,1)$
 &$-$
 &$-$
 &$4.6t_\beta^2\times 10^{-11}$\\
$(7,3,2)$&$(-1,-1,-1)$
 &$-$
 &$-$
 &$-$\\
\hline
$(-1,-1,-1)$&$(7,3,2)$
 &$1.2\times 10^{-10}$
 &$-$
 &$2.1(1+0.060t_\beta)^2\times 10^{-8}$\\
$(1,1,1)$&$(6,2,1)$
 &$1.2(1+0.015t_\beta)^2\times 10^{-10}$
 &$-$
 &$2.1(1+0.37t_\beta)^2\times 10^{-8}$\\
$(2,2,2)$&$(5,1,-1)$
 &$1.3(1+0.092t_\beta)^2\times 10^{-10}$
 & $-$
 &$2.1\times 10^{-8}$\\
$(2,2,2)$&$(5,1,-2)$
 &$1.2(1+0.093t_\beta)^2\times 10^{-10}$
 & $-$
 &$2.1(1+0.015t_\beta)^2\times 10^{-8}$\\
\hline
%\end{tabular}
%\caption{\label{table:lfv1}}
%\end{center}
%\end{table}
%%
%\begin{table}
%\begin{center}
%\begin{tabular}{|c|c|c|c|c|}
\hline
\multicolumn{5}{|c|}{$y^\ell_i={\cal O}(\epsilon^8,\epsilon^3,\epsilon^2)$}\\
\hline
$N^e_i$ & $N^\ell_i$ &
$BR(\mu^+_R \to e^+_L,\gamma)$ &
$BR(\tau^+_R \to e^+_L,\gamma)$ &
$BR(\tau^+_R \to \mu^+_L,\gamma)$
\\
\hline
$(6,1,-2)$&$(2,2,2)$
 &$9.2(1+0.11t_\beta)^2\times 10^{-12}$
 &$-$
 &$2.4(1+0.091t_\beta)^2\times 10^{-8}$\\
$(6,1,-1)$&$(2,2,2)$
 &$9.4(1+0.11t_\beta)^2\times 10^{-12}$
 &$-$
 &$2.1\times 10^{-7}$\\
$(7,2,1)$&$(1,1,1)$
 &$8.2(1+0.021t_\beta)^2\times 10^{-12}$
 &$-$
 &$4.4(1+0.42 t_\beta)^2\times 10^{-8}$\\
$(8,3,2)$&$(-1,-1,-1)$
 &$7.8\times 10^{-12}$
 &$-$
 &$2.4(1+0.091t_\beta)^2\times 10^{-8}$\\
\hline
$(-1,-1,-1)$&$(8,3,2)$
 & $-$
 & $-$
 & $-$\\
$(1,1,1)$&$(7,2,1)$
 & $-$
 & $-$
 & $-$\\
$(2,2,2)$&$(6,1,-1)$
 & $-$
 & $-$
 & $-$\\
$(2,2,2)$&$(6,1,-2)$
 & $-$
 & $-$
 & $-$\\
\hline
$N^e_i$ & $N^\ell_i$ &
$BR(\mu^+_L \to e^+_R,\gamma)$ &
$BR(\tau^+_L \to e^+_R,\gamma)$ &
$BR(\tau^+_L \to \mu^+_R,\gamma)$
\\
\hline
$(6,1,-2)$&$(2,2,2)$
 &$-$
 &$-$
 &$-$\\
$(6,1,-1)$&$(2,2,2)$
 &$-$
 &$-$
 &$-$\\
$(7,2,1)$&$(1,1,1)$
 &$-$
 &$-$
 &$4.6t_\beta^2\times 10^{-11}$\\
$(8,3,2)$&$(-1,-1,-1)$
 &$-$
 &$-$
 &$-$\\
\hline
$(-1,-1,-1)$&$(8,3,2)$
 &$7.7\times 10^{-12}$
 &$-$
 &$2.1(1+0.060t_\beta)^2\times 10^{-8}$\\
$(1,1,1)$&$(7,2,1)$
 &$7.8(1+0.012t_\beta)^2\times 10^{-12}$
 &$-$
 &$2.1(1+0.37t_\beta)^2\times 10^{-8}$\\
$(2,2,2)$&$(6,1,-1)$
 &$7.8(1+0.074t_\beta)^2\times 10^{-12}$
 &$-$
 &$2.1\times 10^{-8}$\\
$(2,2,2)$&$(6,1,-2)$
 &$7.7(1+0.075t_\beta)^2\times 10^{-12}$
 &$-$
 &$2.1(1+0.060t_\beta)^2\times 10^{-8}$\\
\hline
\end{tabular}
\caption{Predictions of lepton flavor violating rates for
 $|\mhalf|=500 {\rm GeV}$.
Here all parameters of ${\cal O}(1)$ are set to 1 and  all leading contributions
 are added constructively, so the actual rates can be somewhat smaller than
the numbers in the table.
The $\mu\to e\gamma$ ($\tau\to e\gamma$ or $\mu \gamma$)
branching ratio smaller than $10^{-14}$ ($10^{-9}$) for
$t_\beta\equiv \tan\beta = 10$ is omitted.
Note that the branching ratios scale as $|\mhalf|^{-4}$.
\label{table:lfv2}}
\end{center}
\end{table}
\begin{table}
\begin{center}
\begin{tabular}{|c|c|c|c|c|}
\hline
\multicolumn{5}{|c|}{$y^\ell_i={\cal O}(\epsilon^8,\epsilon^4,\epsilon^2)$}\\
\hline
$N^e_i$ & $N^\ell_i$ &
$BR(\mu^+_R \to e^+_L,\gamma)$ &
$BR(\tau^+_R \to e^+_L,\gamma)$ &
$BR(\tau^+_R \to \mu^+_L,\gamma)$
\\
\hline
$(6,2,-2)$&$(2,2,2)$
 &$1.3(1+0.024t_\beta)^2\times 10^{-10}$
 &$-$
 &$3.3\times 10^{-9}$\\
$(6,2,-1)$&$(2,2,2)$
 &$1.3(1+0.024t_\beta)^2\times 10^{-10}$
 &$-$
 &$5.0(1+ 0.25t_\beta)^2 \times 10^{-9}$\\
$(7,3,1)$&$(1,1,1)$
 &$1.2\times 10^{-10}$
 &$-$
 &$5.0(1+0.25t_\beta)^2 \times 10^{-9}$\\
$(8,4,2)$&$(-1,-1,-1)$
 &$1.2\times 10^{-10}$
 &$-$
 &$2.6(1+0.047t_\beta)^2 \times 10^{-9}$\\
\hline
$(2,2,-2)$&$(6,2,2)$
 &$3.1\times 10^{-11}$
 &$3.4\times 10^{-9}$
 &$3.3 \times 10^{-9}$\\
\cb{$(2,2,-1)$}&\cb{$(6,2,2)$}
 &\cb{$8.4(1+0.27t_\beta)^2\times 10^{-11}$}
 &\cb{$5.2(1+0.25t_\beta)^2 \times 10^{-9}$}
 &\cb{$5.1(1+0.25t_\beta)^2 \times 10^{-9}$}\\
\cb{$(3,3,1)$}&\cb{$(5,1,1)$}
 &\cb{$8.4(1+0.27t_\beta)^2\times 10^{-11}$}
 &\cb{$5.2(1+0.25t_\beta)^2 \times 10^{-9}$}
 &\cb{$5.1(1+0.25t_\beta)^2 \times 10^{-9}$}\\
\cb{$(4,4,2)$}&\cb{$(4,-1,-1)$}
 &\cb{$3.4(1+0.051t_\beta)^2\times 10^{-11}$}
 &\cb{$3.7(1+0.047t_\beta)^2 \times 10^{-9}$}
 &\cb{$3.6(1+0.047t_\beta)^2 \times 10^{-9}$}\\
$(4,4,2)$&$(4,-2,-2)$
 &$3.4(1+0.053t_\beta)^2\times 10^{-11}$
 &$3.7(1+0.047t_\beta)^2 \times 10^{-9}$
 &$3.6(1+0.047t_\beta)^2 \times 10^{-9}$\\
\hline
$(4,-2,-2)$&$(4,4,2)$
 &$1.9(1+0.12t_{\beta})^2\times 10^{-10}$
 &$-$
 &$-$ \\
\cb{$(4,-1,-1)$}&\cb{$(4,4,2)$}
 &\cb{$3.4(1+0.10t_{\beta})^2\times 10^{-10}$}
 &\cb{$-$}
 &\cb{$-$} \\
\cb{$(5,1,1)$}&\cb{$(3,3,1)$}
 &\cb{$4.9(1+0.12t_{\beta})^2\times 10^{-10}$}
 &\cb{$-$}
 &\cb{$-$} \\
\cb{$(6,2,2)$}&\cb{$(2,2,-1)$}
 &\cb{$2.7(1+0.026t_{\beta})^2\times 10^{-10}$}
 &\cb{$-$}
 &\cb{$-$} \\
$(6,2,2)$&$(2,2,-2)$
 &$1.5(1+0.022t_{\beta})^2\times 10^{-10}$
 &$-$
 &$-$ \\
\hline
$(-1,-1,-1)$&$(8,4,2)$
 &$-$
 &$-$
 &$-$\\
$(1,1,1)$&$(7,3,1)$
 &$-$
 &$-$
 &$-$\\
$(2,2,2)$&$(6,2,-1)$
 &$-$
 &$-$
 &$-$\\
$(2,2,2)$&$(6,2,-2)$
 &$-$
 &$-$
 &$-$\\
\hline
$N^e_i$ & $N^\ell_i$ &
$BR(\mu^+_L \to e^+_R,\gamma)$ &
$BR(\tau^+_L \to e^+_R,\gamma)$ &
$BR(\tau^+_L \to \mu^+_R,\gamma)$
\\
\hline
$(6,2,-1)$&$(2,2,2)$
 &$-$
 &$-$
 &$-$\\
$(6,2,-1)$&$(2,2,2)$
 &$-$
 &$-$
 &$-$\\
$(7,3,1)$&$(1,1,1)$
 &$-$
 &$-$
 &$-$\\
$(8,4,2)$&$(-1,-1,-1)$
 &$-$
 &$-$
 &$-$\\
\hline
$(2,2,-2)$&$(6,2,2)$
 &$1.3(1+0.014t_\beta)^2  \times 10^{-10}$
 &$-$
 &$-$\\
\cb{$(2,2,-1)$}&\cb{$(6,2,2)$}
 &\cb{$5.0(1+0.015t_\beta)^2  \times 10^{-10}$}
 &\cb{$-$}
 &\cb{$-$}\\
\cb{$(3,3,1)$}&\cb{$(5,1,1)$}
 &\cb{$5.8 (1+0.082t_\beta)^2 \times 10^{-10}$}
 &\cb{$-$}
%$5.2 t_\beta^2 \times  10^{-12}$
 &\cb{$-$}\\
\cb{$(4,4,2)$}&\cb{$(4,-1,-1)$}
 &\cb{$2.3(1+0.079t_{\beta})^2\times 10^{-10}$}
 & \cb{$-$}
 &\cb{$-$}\\
$(4,4,2)$&$(4,-2,-2)$
 &$1.8(1+0.32t_{\beta})^2\times 10^{-10}$
 &$-$
 &$-$\\
\hline
$(4,-2,-2)$&$(4,4,2)$
 &$3.1(1+0.032 t_\beta)^2\times 10^{-11}$
 &$3.4(1+0.030t_\beta)^2 \times 10^{-9}$
 &$3.3(1+0.030 t_\beta)^2 \times 10^{-9}$ \\
\cb{$(4,-1,-1)$}&\cb{$(4,4,2)$}
 &\cb{$3.1(1+0.032 t_\beta)^2\times 10^{-11}$}
 &\cb{$3.4(1+0.030t_\beta)^2 \times 10^{-9}$}
 &\cb{$3.3(1+0.030 t_\beta)^2 \times 10^{-9}$} \\
\cb{$(5,1,1)$}&\cb{$(3,3,1)$}
 &\cb{$3.1(1+0.20 t_\beta)^2\times 10^{-11}$}
 &\cb{$3.4(1+0.19t_\beta)^2 \times 10^{-9}$}
 &\cb{$3.3(1+0.19t_\beta)^2 \times 10^{-9}$} \\
\cb{$(6,2,2)$}&\cb{$(2,2,-1)$}
 &\cb{$3.1(1+0.20t_\beta)^2\times 10^{-11}$}
 &\cb{$3.4(1+0.19t_\beta)^2 \times 10^{-9}$}
 &\cb{$3.3(1+0.19t_\beta)^2 \times 10^{-9}$} \\
$(6,2,2)$&$(2,2,-2)$
 &$3.1\times 10^{-11}$
 &$3.4\times 10^{-9}$
 &$3.3\times 10^{-9}$ \\
\hline
$(-1,-1,-1)$&$(8,4,2)$
 &$1.2\times 10^{-10}$
 &$-$
 &$3.3(1+0.030t_\beta)^2 \times 10^{-9}$\\
$(1,1,1)$&$(7,3,1)$
 &$1.2\times 10^{-10}$
 &$-$
 &$3.3(1+0.19t_\beta)^2\times 10^{-9}$\\
$(2,2,2)$&$(6,2,-1)$
 &$1.2(1+0.015t_\beta)^2\times 10^{-10}$
 &$-$
 &$3.3(1+0.19t_\beta)^2\times 10^{-9}$\\
$(2,2,2)$&$(6,2,-2)$
 &$1.2(1+0.015t_\beta)^2\times 10^{-10}$
 &$-$
 &$3.3\times 10^{-9}$\\
\hline
\end{tabular}
\caption{Predictions of lepton flavor violating rates.
Models indicated by light color
can not satisfy the current experimental bound on $\mu \to e\gamma$
even when the involved ${\cal O}(1)$ parameters are assumed to be
suppressed by a factor of $1/4$
or $|\mhalf|=1$ TeV.
\label{table:lfv4}}
\end{center}
\end{table}

Let us now examine the quark sector. It is well known that
CP violating parameter $\epsilon_K$ in $K$-$\overline{K}$ mixing
 is quite sensitive to the supersymmetric extension of the standard model.
CP is conserved in the SM if there is no third generation.
Consequently, the SM contribution to
$\epsilon_K$ is suppressed by small CKM mixing angles
compared to naive dimensional estimation.
However, this is not the case for the supersymmetric models,
thus the SUSY contribution to $\epsilon_K$ can easily become comparable
to the SM contribution.
The gluino mediated contribution to $\epsilon_K$,
normalized by the experimental value \cite{Hagiwara:fs}, can be summarized as
\bea
\left[\frac{\epsilon_K}{2.282 \times10^{-3}}\right]
&\simeq&
e^{i\frac{\pi}{4}}
\left(
\frac{500 {\rm GeV}}{|\mhalf|}
\right)^2
%\nonumber\\
%&&
%\times
\left[
\frac{ {\rm Im}[(\delta^d_{LL})^2_{12}]}{(1.5\times 10^{-2})^2}
+\frac{{\rm Im}[(\delta^d_{RR})^2_{12}]}{(1.5\times 10^{-2})^2}
%\right.
%\nonumber\\
%&&\phantom{\times[}
-\frac{{\rm Im}[(\delta^d_{RR})_{12}(\delta^d_{LL})_{12}]}
{(2.2\times 10^{-4})^2}
\right.
\nonumber\\
&&
\phantom{
e^{i\frac{\pi}{4}}
\left(
\frac{500 {\rm GeV}}{|\mhalf|}
\right)^2
}
\phantom{\times[}
\left.
+\frac{{\rm Im}[(\delta^d_{RL})^2_{12}]
+{\rm Im}[(\delta^d_{LR})^2_{12}]}
{(0.63\times 10^{-3})^2}
%\nonumber\\
%&&\left.\phantom{\times[}
-\frac{{\rm Im}[(\delta^d_{RL})_{12}(\delta^d_{LR})_{12}]}
{(0.49\times 10^{-3})^2}
\right],
\label{eq:epsilon_K}
\eea
where we assume again  the sparticle mass spectrum
of (\ref{rg}).
% and allow up to two off-diagonal mass-insertions.
%Generalized formula is given in appendix \ref{sec:mass_insertion}
Analytic formulas for the corresponding Wilson coefficients can be
 found in \cite{massinsertion}.
Here we followed \cite{Buras:2001ra} to estimate the QCD corrections and relevant
hadronic matrix elements, and Table \ref{table:hadronic_constants} for the involved phenomenological
numbers.
\begin{table}[h]
\begin{center}
\begin{tabular}{|c|c|c|c|c|c|c|c|}
\hline
$\alpha_s(M_Z)$ & $M_K$   & $\Delta M_K$ & $F_K$   & $M_{B_d}$ & $F_{B_d}$ & $M_{B_s}$
& $F_{B_s}/F_{B_d}$ \\
\hline
0.119 & 494 MeV & $5.30 ns^{-1} \hbar$ & 160 MeV & 5.28 GeV  & 195 MeV   &  5.37 GeV
& 1.21 \\
\hline
\end{tabular}
\caption{Input parameters used in the mass-insertion formulas. For the
 hadronic matrix elememts, we follow Ref.~\cite{Buras:2001ra}.  \label{table:hadronic_constants}}
\end{center}
\end{table}
The typical size of
 $(\delta^d_{RL})_{12,21}$ in our SS-mediated SUSY breaking models
 is ${\cal O}(10^{-4})$
as can be seen from the expression of $\tilde{\delta}^d_{RL}$
analogous to (\ref{eq:deltalG_RL12}) and also (\ref{rg}),
therefore the RL insertions
do not give an observable contribution to $\epsilon_K$.
Also the contributions from
${\rm Im}[(\delta^d_{LL,RR})^2_{12}]$ are
relatively small because the relevant matrix element
does not receive  QCD enhancement.
In our case, accidental cancellation of relevant mass functions
reduces these contributions further.
Then the most dominant contribution comes from
${\rm Im}[(\delta^d_{RR})_{12}(\delta^d_{LL})_{12}]$.
Higher order insertions of $\delta$
including the third generation are not important
because  (\ref{eq:deltad_RR23})
 and (\ref{rg}) show
 $|(\delta^d_{RR})_{13}(\delta^d_{RR})_{32}| \lsim
 |(\delta^d_{RR})_{12}|$ and a similar relation for $(\delta^d_{LL})_{12}$.

Upper half of the Table \ref{table:epsilon_K} summarizes the resulting
$\epsilon_K$ for
the effective flavor charges of (\ref{favored}) providing a best fit
to the observed  CKM matrix and down-type quark masses
under the assumption that $\kappa_{q,d}$ and
 $e^{i(\phi^d_1-\phi^d_2)}$ are complex in general.
The results are expressed in terms of
the $\delta$ parameters normalized by their  values saturating
 the observed $|\epsilon_K|$.
Again here we choose $|\mhalf|\sim 500$ GeV and assume
that the complex parameters $\kappa_{q,d}$ and the phases
$\phi^d_i$ are all of order unity.
For these four models, the contribution
from ${\rm Im}[(\delta^d_{RR})_{12}(\delta^d_{LL})_{12}]$
exceeds the observed value by one or two orders of magnitude,
thus one needs a fine tuning of ${\cal
 O}(10^{-1}\sim 10^{-2})$ for the involved parameters.
Even when $\kappa_{q,d}$ and
 $e^{i(\phi^d_1-\phi^d_2)}$ are all real, the situation is not improved
much.
In the presence of the KM phase, the RG evolution from $M_c$ to $M_Z$
generates a CP violating phase in $(\delta^d_{LL})_{12}$:
\bea
{\rm Im}[(\delta^d_{LL})_{12}] &\sim& \frac{1}
{\sqrt{m^{2(\tilde{q})}_{11}m^{2(\tilde{q})}_{22}}}
{\rm Im}[(A_u^\dag A_u)_{12}]\frac{1}{(4\pi)^2} \ln \left(
\frac{M_Z^2}{M_c^2}\right)
\nonumber\\
&\sim& -6\times 10^{-2} \left(\frac{|\mhalf|}{v_u}\right)^2
{\rm Im}[(\tilde{\delta}^{u}_{RL})^*_{31}(\tilde{\delta}^{u}_{RL})_{32}]
\nonumber\\
&\sim& -10^{-3}\, {\rm Im}[(V_{CKM}\bar{V}_q)_{13}(V_{CKM}\bar{V}_q)_{23}^*]
/\epsilon^{X^q_1+X^q_2-2X_3^q},
\label{eq:RGE effect}
\eea
where $v_u =\langle H_2^0\rangle \simeq 174 \sin\beta$ GeV.
This RG induced contribution
is numerically only a factor few smaller than
the direct contribution at $M_c$ coming from complex $\kappa_{q,d}$
and $e^{i(\phi^d_1-\phi^d_2)}$.

If we relax the condition (\ref{favored}) for the best fit to
the quark masses and CKM matrix, we can choose
$N^d_1=N^d_2=N^d_3$ for the down-type
quark singlets, which would make all $(\delta^d_{RR})_{ij}$ disappear.
Note that the SS-induced $m^{2(\tilde{d})}_{i\bar{j}}$ at the compactification scale
are universal if $N^d_i$ are all degenerate.
Lower half of the Table
\ref{table:epsilon_K} shows that in this case
there is no observable deviation of $\epsilon_K$.
However in this case, in order to produce the correct quark masses and CKM matrix,
we have to assume that some boundary Yukawa couplings
are abnormally large (or small) by a factor of $4\sim 5$ ($0.2\sim 0.3$)
compared to the values suggested by the naive dimensional analysis \cite{luty}.
For instance,
the model with
$X^d_i=(3,2,0)$ requires that
the boundary Yukawa coupling $\tilde{\lambda}_{11}$ is smaller than the
naively expected value by a factor of $0.2$.
%On the other hands, the model with $X^d_i=(4,3,0)$
% $(V_{CKM})_{23,32}\simeq {\cal O}(\epsilon^3)$/
%$(V_{CKM})_{13,31}\simeq {\cal O}(\epsilon^4)$ for $-N^d_i=(4,3,0)$
%and $(V_{CKM})_{12,21}\simeq {\cal O}(\epsilon^2)$/
%$(V_{CKM})_{13,31}\simeq {\cal O}(\epsilon^4)$ for $-N^d_i=(4,2,0)$.
%In this case, an attractive possibility is
% bulk SU(5) unification of $\ell$ and $d$ in $\overline{\bf 5}$
% that naturally leads to the universal $X^d$
% from the observed neutrino mixings,
% however, it is hard
% to reconcile the Cabibbo angle ($X^q$) and $\mu \to
% e,\gamma$ ($X^e$) in $\bf 10$ without further assumption
% on the Yukawa texture.

\begin{table}
\begin{center}
\begin{tabular}{|c|c|c|c|c|}
\hline
$N^q_i$ & $N^d_i$ &
$\frac{{\rm Im}[(\delta^d_{LL})^2_{12}]}{(1.5\times 10^{-2})^2}$ &
$\frac{{\rm Im}[(\delta^d_{RR})^2_{12}]}{(1.5\times 10^{-2})^2}$ &
$\frac{{\rm Im}[(\delta^d_{RR})_{12}(\delta^d_{LL})_{12}]}{(2.2\times 10^{-4})^2}$ \\
\hline
$(3,2,-1)$ & $(3,2,2)$ &$1.5\times 10^{-2}$ &$7.1\times 10^{-1}$ & $466$\\
$(3,2,-1)$ & $(4,2,2)$ &$1.5\times 10^{-2}$ &$2.8\times 10^{-2}$ & $93$\\
\hline
$(3,2,-1)$ & $(3,3,2)$ &$1.5\times 10^{-2}$ &$2.8\times 10^{-2}$ & $93$\\
$(3,2,-1)$ & $(4,3,2)$ &$1.5\times 10^{-2}$ &$5.7\times 10^{-3}$ & $42$\\
\hline
\hline
$(3,2,-1)$ & $(2,2,2)$ &$1.5\times 10^{-2}$ &0& $0$ \\
$(4,3,-1)$ & $(2,2,2)$ &$1.2\times 10^{-4}$ &0& $0$ \\
$(4,2,-1)$ & $(2,2,2)$ &$6.0\times 10^{-4}$ &0& $0$ \\
\hline
\end{tabular}
\caption{Quark mass hierarchy vs $\epsilon_K$.
Here $\delta^d$ 's are divided by the values  saturating
$\epsilon_K = 2.282 \times 10^{-3}$
 for $|\mhalf|=500 {\rm GeV}$.
%Models in the upper half fit well the observed quark masses and CKM matrix
%elements with the boundary Yukawa couplings suggested by the naive dimensional analysis.
%However for the models in the lower half, one needs to assume that some boundary Yukawa couplings
%a factor $4\sim 5$ ($0.2 \sim 0.3$)
 %tuning of ${\cal O}(1)$ parameters in the fit.
\label{table:epsilon_K}}
\end{center}
\end{table}

For the sparticle spectrum (\ref{rg}), the gluino contribution to
the $K^0$-$\overline{K}^0$ and $B_{d,s}^0$-$\overline{B}_{d,s}^0$ mass differences
in SS SUSY breaking scenario
are given by
\begin{eqnarray}
\left[
%\frac{\Delta M_K}{3.49\times 10^{-15} {\rm GeV}}\right]
\frac{\Delta M_K}{5.30 {\rm ns^{-1}}\hbar}\right]
&\simeq&
\left(
\frac{500 {\rm GeV}}{|\mhalf|}
\right)^2
%\nonumber\\
%&&
%\times
\left[
\frac{ {\rm Re}[(\delta^d_{LL})^2_{12}]}{(1.8\times 10^{-1})^2}
+\frac{{\rm Re}[(\delta^d_{RR})^2_{12}]}{(1.9\times 10^{-1})^2}
%\right.
%\nonumber\\
%&&\phantom{\times[}
-\frac{{\rm Re}[(\delta^d_{RR})_{12}(\delta^d_{LL})_{12}]}
{(2.8\times 10^{-3})^2}
\right.
\nonumber\\
&&\phantom{\times[}
\phantom{
\left(
\frac{500 {\rm GeV}}{|\mhalf|}
\right)^2
}
\left.
+\frac{{\rm Re}[(\delta^d_{RL})^2_{12}]
+{\rm Re}[(\delta^d_{LR})^2_{12}]}
{(0.78\times 10^{-2})^2}
%\nonumber\\
%&&
%\left.
%\phantom{\times[}
-\frac{{\rm Re}[(\delta^d_{RL})_{12}(\delta^d_{LR})_{12}]}
{(0.60\times 10^{-2})^2}
\right],
\nonumber\\
%\end{eqnarray}
%%
%\begin{eqnarray}
\left[\frac{\Delta M_{B_d}}{0.489 {\rm ps^{-1}}\hbar}\right]
&\simeq&
\left(
\frac{500 {\rm GeV}}{|\mhalf|}
\right)^2
%\nonumber\\
%&&
%\times
\left|
\frac{(\delta^d_{LL})^2_{13}}{(3.5\times 10^{-1})^2}
+\frac{(\delta^d_{RR})^2_{13}}{(3.4\times 10^{-1})^2}
%\right.
%\nonumber\\
%&&
%\phantom{\times[}
-\frac{(\delta^d_{RR})_{13}(\delta^d_{LL})_{13}}
{(3.2\times 10^{-2})^2}
\right.
\nonumber\\
&&
\phantom{
\left(
\frac{500 {\rm GeV}}{|\mhalf|}
\right)^2
}
\phantom{\times[}
\left.
+\frac{(\delta^d_{RL})^2_{13}}{(0.65\times 10^{-1})^2}
+\frac{(\delta^d_{LR})^2_{13}}
{(0.66\times 10^{-1})^2}
%\nonumber\\
%&&
%\left.
%\phantom{\times[}
-\frac{(\delta^d_{RL})_{13}(\delta^d_{LR})_{13}}
{(0.71\times 10^{-1})^2}
\right|,
\nonumber\\
%\end{eqnarray}
%%
%\begin{eqnarray}
\left[\frac{\Delta M_{B_s}}{13.1 {\rm ps^{-1}}\hbar}\right]
&\simeq&
\left(
\frac{500 {\rm GeV}}{|\mhalf|}
\right)^2
%\nonumber\\
%&&
%\times
\left|
\frac{(\delta^d_{LL})^2_{23}+(\delta^d_{RR})^2_{23}}{(1.5)^2}
%\right.
%\nonumber\\
%&&
%\phantom{\times[}
-\frac{(\delta^d_{RR})_{23}(\delta^d_{LL})_{23}}
{(1.4\times 10^{-1})^2}
\right.
\nonumber\\
&&
\phantom{
\left(
\frac{500 {\rm GeV}}{|\mhalf|}
\right)^2
}
\phantom{\times[}
\left.
+\frac{(\delta^d_{RL})^2_{23}+(\delta^d_{LR})^2_{23}}
{(2.8\times 10^{-1})^2}
%\nonumber\\
%&&\left.
%\phantom{\times[}
-\frac{(\delta^d_{RL})_{23}(\delta^d_{LR})_{23}}
{(3.0\times 10^{-1})^2}
\right|.
\label{massdif}
\end{eqnarray}
Here we followed
Ref.\cite{massinsertion}
for the involved Wilson coefficients,
Ref.\cite{Buras:2001ra} for the
 QCD corrections and hadronic matrix elements, and Table \ref{table:hadronic_constants}
for the involved phenomenological numbers.
The results are then well below
the experimental values ($\Delta M_{K,B_d}$) or the latest
upper bound ($\Delta M_{B_s}$) \cite{Hagiwara:fs},
typically less than 1 \%.
%General formulas are given in the appendix \ref{}.
In fact, for $\Delta M_{B_d}$,
there can be few \% SUSY contributions
which were not included in (\ref{massdif}) as they
come from the RG effects involving the top quark Yukawa coupling
similarly to the effects of (\ref{eq:RGE effect}).
%which may be explored by precision measurements at B-factories and their successors
%\cite{Bigi:2004kn, Ball:2000ba}.
% along with progress of non--perturbative QCD calculation.
%Of course, for those four models,
%we need to assume a fine tuning of
%${\cal O}(10^{-1})$ for CP
%violating parameters to make $\epsilon_K$ small enough.
%Whereas, for the lower three models
%which give negligible SUSY contribution
%to $\epsilon_K$,
%the SUSY contributions to $\Delta M_{B_{d,s}}$ are also well
% below the experimental reach.
Flavor violating soft parameters can affect also
$\epsilon'/\epsilon_K$ even when they do not contain any new CP violating
phase. We have estimated the  gluino
 contribution to $\epsilon'/\epsilon_K$
in SS SUSY breaking scenario, and find that it is at most comparable to
the SM contribution for a reasonable range of the involved parameters.
Because the consensus on the SM contribution to $\epsilon'/\epsilon_K$
 has not been achieved
yet \cite{epsilon'}, we can not derive any  meaningful constraint from this result.

Let us finally consider the SUSY
 contribution to $b \to s \gamma$ in our models.
The branching ratio can be approximated by
\begin{equation}
BR[\overline{B} \to X_s, \gamma]^{\rm SUSY} \simeq
\Delta BR(b_R \to s_L, \gamma)^{int}+BR(b_L \to s_R, \gamma),\\
\end{equation}
where the first term denotes interference with the SM contribution
 and the second term comes from the operators of opposite chirality
 to the SM ones.
The gluino contributions normalized by the latest
 world average \cite{bsgamma_exp} are given by
\begin{eqnarray}
\left[
\frac{\Delta BR(b_R \to s_L, \gamma)^{int}}{3.34\,(\pm 0.38)\times 10^{-4}}
\right]
&\simeq&
\left(\frac{500 {\rm GeV}}{|\mhalf|}\right)%^2
\left.
\left\{
\frac{{\rm Re}[(\delta^d_{LR})_{23}]}{0.021}
-\frac{{\rm Re}[(\delta^d_{LL})_{21}(\delta^d_{LR})_{13}]}{0.040}
-\frac{{\rm Re}[(\delta^d_{LL})_{23}(\delta^d_{LR})_{33}]}{0.043}
\right.\right.\nonumber\\
&&
\left.
\phantom{
\left(\frac{500 {\rm GeV}}{|\mhalf|}\right)\{
}
-\frac{{\rm Re}[(\delta^d_{LR})_{21}(\delta^d_{RR})_{13}]+{\rm Re}[(\delta^d_{LR})_{22}(\delta^d_{RR})_{23}]}{0.041}
\right\}
%\right]
,\\
\left[
\frac{BR(b_L \to s_R, \gamma)}{3.34\,(\pm 0.38)\times 10^{-4}}
\right]^{1/2}
&\simeq&
\left(\frac{500 {\rm GeV}}{|\mhalf|}\right)%^2
%\nonumber\\
%&&\times
%\left|
%\frac{(\delta_{RR})_{23}}{1.5\times 10}
%-\frac{(\delta^d_{RR})_{21}(\delta^d_{RR})_{13}}{2.7\times 10^{2}}
%\right.\nonumber\\
%&&\left.\phantom{\times|}
%+
%\left(
%\frac{|\mhalf|}{500 {\rm GeV}}
%\right)
%\left(
%\frac{2.71 {\rm GeV}}{\overline{m}_b(m_t)}
%\right)
%\right.
%\nonumber\\
%&&
%\times
\left|
%\left.
%\phantom{\times[+}\times
%\left\{
-\frac{(\delta^d_{RL})_{23}}{0.042}
+\frac{(\delta^d_{RR})_{21}(\delta^d_{RL})_{13}+(\delta^d_{RR})_{23}(\delta^d_{RL})_{33}}{0.082}
%\right.
\right.\nonumber\\
&&
%\left.
\left.
%\phantom{\times[+}%\phantom{+\times~}
\phantom{
\left(\frac{500 {\rm GeV}}{|\mhalf|}\right)|
}
+\frac{(\delta^d_{RL})_{21}(\delta^d_{LL})_{13}+(\delta^d_{RL})_{22}(\delta^d_{LL})_{23}}{0.080}
%\right\}
\right|,
\end{eqnarray}
for the sparticle spectrum (\ref{rg}).
Here we followed Ref.\cite{bsgamma} to
estimate the branching ratio from the relevant Wilson coefficients at
$m_t$ and included the leading--order gluino contribution to
$C_{4,7,8}$ and their chirality partners.
The Ref.\cite{bsgamma}
 quotes the SM contribution as
 $BR[\overline{B} \to X_s\gamma]_{E_0>\frac{m_b}{20}} = 3.70\pm 0.30
 \times 10^{-4}$. This overlaps with the experimental value within
 the 1 $\sigma$ errors $\sim 10\%$.
In our models, $|\delta^d_{RL,LR}|$ is at most $10^{-3}$.
Therefore their contributions
 to the interference term is only a few \% level.
A potentially dangerous contribution may come from
$(\delta^d_{LL})_{23}(\delta^d_{LR})_{33}$
associated  with the F-term contribution to
 $(\delta^d_{LR})_{33}$
which can be estimated
as
$$(\delta^d_{LR})_{33}\simeq -1.5\times 10^{-3} e^{i\theta_\mu} \tan\beta
 (500 {\rm GeV}/|\mhalf|)$$
for a moderate value $\tan\beta$.
For the models listed in Table \ref{table:epsilon_K},
$(\delta^d_{LL})$ is at most a few \% for $|\mhalf| \sim  500$ GeV,
and then its contribution to $b \to s\gamma$ is also below a few \%.
The pure SUSY contribution from the opposite chirality is even more
negligible because it scales quadratically with the SUSY amplitude,
so typically gives a correction of ${\cal O}(10^{-3})$.
Consequently the gluino mediated contribution to $b \to s\gamma$
 does not give any meaningful constraint for our models with the current
 experimental and theoretical accuracy.

\section{Conclusion}

Quasi-localization of matter fields in extra dimension
is an elegant mechanism to generate hierarchical 4D
Yukawa couplings.
Extra dimension provides also an attractive way
to break SUSY by boundary conditions as originally proposed
by Scherk and Schwarz.
In this paper, we have examined some physical consequences
of implementing the quasi-localization of matter zero modes
and the SS SUSY breaking
simultaneously within 5D orbifold field theories.
In this case, the radion corresponds to
a flavon to generate the flavor hierarchy and at the same time
plays the role of the messenger of supersymmetry breaking.
As a consequence, the resulting soft scalar masses and
trilinear $A$-parameters of matter zero modes
at the compactification scale are highly flavor-dependent,
thereby can lead to dangerous flavor violations at low energy scales.
The shape of soft parameters implies also that the compactification scale
should be much higher than the weak scale in order for the model
to be phenomenologically viable.

We have computed the soft parameters of quasi-localized matter fields
induced by the SS boundary condition in generic
5D orbifold SUGRA.
It is shown explicitly that the zero mode soft parameters from
the SS boundary condition are same as the ones
induced by the radion $F$-component in 4D effective SUGRA,
and thus our analysis applies to any SUSY breaking mechanism
giving a sizable $F$-component of the radion superfield,
e.g. the hidden gaugino condensation model.

In 5D orbifold SUGRA, quasi-localization of matter zero modes are
governed by the kink masses of matter hypermultiplets,
$M_I\epsilon(y)$, which
have quantized-values if the graviphoton and/or $U(1)_{FI}$ gauge charges are quantized.
An important feature of the SS SUSY breaking or the radion-mediated SUSY breaking
is that,
if the kink masses are quantized,
the resulting soft scalar masses and the trilinear scalar couplngs (divided by the corresponding
Yukawa couplings) at the compactification scale are quantized also in the leading approximation.
This feature provides a natural mechanism to suppress dangerous flavor violations
since the flavor violating amplitudes appear in a form
$f(M_I)-f(M_J)$, thus are canceled
when some of the quantized kink masses are degenerate.

We analyzed in detail the low energy flavor violations
in SS-dominated supersymmetry breaking scenario
under the assumption that  the compactification scale $M_c$
is close to the grand unification scale $\sim 2\times 10^{16}$ GeV
and the 4D effective theory below $M_c$ is the minimal
supersymmetric standard model.
We find that many of the low energy flavor violations are appropriately
suppressed,
however generically $\epsilon_K$ and $\mu\rightarrow e\gamma$
can be dangerous if the SS boundary condition is the major source of
SUSY breaking.
Assuming that the hypermultiplet kink masses
are quantized,
the $\mu\rightarrow e\gamma$ bound can be satisfied
for a reasonable range of the involved continuous parameters
if either the $SU(2)_L$ doublet lepton kink masses or
the $SU(2)_L$ singlet lepton kink masses are flavor-independent.
These two possibilities are  clearly distinguished by the predicted
chirality pattern of the lepton flavor violating decays.
The chirality structure for degenerate $SU(2)_L$-doublet lepton kink masses
is opposite to the other case with degenerate $SU(2)_L$-singlet lepton kink masses
which has the same chirality structure as
the lepton flavor violating decays induced by the right-handed neutrino Yukawa couplings
in seesaw models \cite{seesaw,Hisano:1998fj,muegpol}.
The SUSY contribution to $\epsilon_K$ can be similarly suppressed
by choosing the quantized kink masses of down-type quarks
to be degenerate. However in this case,
to get the correct quark mass spectrum and CKM mixing angles, one needs to assume
that some boundary Yukawa couplings are abnormally large (or small)
by a factor of $4\sim 5$ ($0.2\sim 0.3$) compared to the values
suggested by the naive dimensional analysis.

\bigskip

{\bf Acknowledgements}

\medskip

This work is supported by KRF PBRG 2002-070-C00022.

\end{document}